\documentclass[ALICE,manyauthors]{cernphprep}
\newcommand{\jpsi}{\ensuremath{\mathrm{J}/\psi}}

\newcommand{\pb}{Pb-Pb}
\newcommand{\pt}{$p_{\rm T}$}
\newcommand{\snn}{$\sqrt{s_\mathrm{NN}}$}

\newcommand{\acceps}{\ensuremath{(\mathrm{Acc}\times\varepsilon)_{\mathrm{J}/\psi}}}
\newcommand{\minv}{$M_\mathrm{inv}$}
\usepackage{rotating}
\usepackage{lineno}
\usepackage{color}
\usepackage{epsfig}
\usepackage{hyperref,cite}
\begin{document}%
%
%
\begin{titlepage}
\PHnumber{2013-066}
\PHdate{April 9, 2013}
%
%
\title{Charmonium and $\mathbf{\it{e}^+\it{e}^-}$ pair photoproduction 
       at mid-rapidity in ultra-peripheral Pb-Pb collisions at $\mathbf{\sqrt{{\it s}_{\rm NN}}}$=2.76 TeV}
\ShortTitle{Charmonium and $e^+e^-$ pair photoproduction}   
%
%
%
%
\Collaboration{ALICE Collaboration%
         \thanks{See Appendix~\ref{app:collab} for the list of collaboration
                      members}}
\ShortAuthor{ALICE Collaboration}      

\begin{abstract}
The ALICE Collaboration at the LHC has
measured the \jpsi~and $\psi^{'}$ photoproduction at mid-rapidity in ultra-peripheral \pb-collisions at \snn=2.76 TeV.
The charmonium is identified via its leptonic decay for events where the hadronic activity is
required to be minimal.
The analysis is based on an event sample corresponding to an integrated luminosity
of about 23 $\mu\rm{b}^{-1}$.
The cross section for coherent and incoherent 
\jpsi~production in the rapidity interval -0.9~$<$$y$$<$~0.9, 
are $\mathrm{d}\sigma_{J/\psi}^{coh}/\mathrm{d}y =
2.38^{+0.34}_{-0.24}\big(\mathrm{sta+sys}\big)$~mb.
and
$\mathrm{d}\sigma_{J/\psi}^{inc}/\mathrm{d}y = 
0.98^{+0.19}_{-0.17}\big(\mathrm{sta+sys}\big)$~mb,
respectively.
The results are compared to theoretical models for \jpsi~production and the coherent cross section is found to be in
good agreement with those models incorporating moderate nuclear gluon shadowing, such as EPS09 parametrization. 
In addition the cross section for
the process $\gamma\gamma\rightarrow e^+ e^-$ has been measured
and found to be in agreement with models implementing QED at leading order.
\end{abstract}
\end{titlepage}
\setcounter{page}{2}

\section{Introduction}
The strong electromagnetic fields generated by heavy ions at the LHC provide 
an opportunity to study photonuclear interactions in ultra-peripheral collisions (UPC), where
the impact parameter may be several tens of femtometres and no hadronic interactions occur. 
The photon flux is proportional to the square of the nucleus charge, so the photon flux in 
lead beams is enhanced by nearly four orders of magnitude compared to 
proton beams.
The strong photon flux leads to large cross sections for a variety of photonuclear and 
two-photon interactions. The physics 
of ultra-peripheral collisions is described in Refs.~\cite{Review2008,Review2005}. 
Exclusive vector meson photoproduction, 
where a vector meson is produced in an event with no 
other final state particles, 
is of particular interest, since it provides a measure of the nuclear gluon distribution at low Bjorken-$x$.

Exclusive production of charmonium in photon-proton interactions at HERA,
$\gamma+ \mathrm{p} \rightarrow \jpsi(\psi^{'}) + \mathrm{p}$, has been
successfully modelled in perturbative QCD in terms of the exchange of two gluons with no net-colour
transfer\ \cite{Frankfurt:1997fj}.
Exclusive vector meson production at mid-rapidity in heavy-ion collisions has previously been 
studied at RHIC~\cite{Abelev:2007nb,Afanasiev:2009hy}.
The exclusive photoproduction can be either coherent, where the
photon couples coherently to almost all the nucleons, or incoherent, where the photon couples to a single
nucleon. Coherent production is characterized by low transverse momentum of vector mesons
($\langle p_{\rm T} \rangle \approx$~60~MeV/$c$) where the nucleus normally does not break up by the \jpsi~production. 
However the exchange of aditional photons may lead to the nucleus break-up, estimated by the simulation models at the level 
of 20-30$\%$ of the events.
Incoherent production,
corresponding to quasi-elastic scattering off a single nucleon, is characterized by a somewhat higher
transverse momentum ($\langle p_{\rm T} \rangle \approx$~500~MeV/$c$). In this case the nucleus interacting with the photon breaks up, but,
apart from single nucleons or nuclear fragments in the very forward region, no other particles are
produced. 

Recently the ALICE Collaboration
published the first results on the photoproduction of \jpsi~in ultra-peripheral Pb-Pb collisions at the LHC~\cite{aliceupcmuonarm}.
This first measurement was performed in the rapidity region\\ \mbox{--3.6~$< y <$--2.6} and allows us to constrain the 
nuclear gluon distribution at Bjorken-$x$~$\approx 10^{-2}$.
In this paper, results from the ALICE experiment on exclusive photoproduction of \jpsi~and $\psi^{'}$~ 
mesons at mid-rapidity in ultra-peripheral Pb-Pb collisions at $\sqrt{s_{\mathrm{NN}} } = 2.76$ TeV are presented. 
 The measurement at mid-rapidity allows the exploration of 
the region $x = (M_{\jpsi}/\sqrt{s_\mathrm{NN}})\rm{exp}(\pm~\it{y})~\approx \rm{10^{-3}}$,
where at present the uncertainty in the nuclear gluon shadowing distribution is rather large\ \cite{Eskola:2009uj}. 
This analysis is focused both on coherently and incoherently produced {\jpsi}~mesons. 
The measured cross section is compared to model 
predictions~\cite{Rebyakova:2011vf,starlight,Adeluyi:2012ph,Goncalves:2011vf,Cisek:2012yt,LM}.

Two-photon production of di-lepton pairs in heavy-ion interactions is also of great interest, 
as it probes Quantum Electrodynamics in the regime of strong fields. 
This process is sensitive to the effect produced by the strong fields of the nuclei. 
The coupling Z$\sqrt{\alpha}$ is large, so higher-order terms may become important. 
Predictions exist where these terms are found 
to lead to a reduction in the cross section by up to 30$\%$~\cite{Baltz:2009fs}. 
Other calculations have found agreement with leading-order calculations for muon pairs 
and electron pairs with invariant masses much larger than two times the electron mass~\cite{Hencken:2006ir}.
Measurements at LHC energies can provide useful insights to assess these effects. 
In this paper we present the study of the $\gamma \gamma \rightarrow e^{+}e^{-}$ process. The results are compared with predictions by models 
neglecting higher order effects discussed above.

\section{Detector description} 
The ALICE experiment consists of a central barrel placed in a large solenoid magnet (B = 0.5 T),
covering the pseudorapidity region $\vert\eta\vert$ $<$ 0.9, and a muon
spectrometer at forward rapidity, covering the range\newline 
\hbox{--4.0~$<\eta<$~--2.5~\cite{Aamodt:2008zz}}. 
In this analysis the following detectors of the central barrel have been used.
The Silicon Pixel Detector (SPD) makes up the two innermost layers of the ALICE Inner Tracking System (ITS), covering extended pseudorapidity ranges
$|\eta| < 2$ and $|\eta| < 1.4$, for the inner (radius 3.9 cm)  and outer (average radius 7.6 cm) layers, respectively. It is a fine granularity detector, having about $10^{7}$ pixels, and can be used for triggering purposes. The Time Projection Chamber (TPC) is used for tracking and for particle identification. A 100 kV central electrode separates 
the two drift volumes, providing an electric field for electron drift,  and the two end-plates, at $|z|= 250$ cm, are instrumented with Multi-Wire-Proportional-Chambers (MWPCs) with 560 000 readout pads, allowing high precision track measurements in the transverse plane. The $z$~coordinate is given by the drift time in the TPC electric field.  The TPC acceptance covers the pseudorapidity region $|\eta| < 0.9$. 
Ionization measurements made along track clusters are used for particle identification\cite{TPCion}. Beyond the TPC, the Time-of-Flight detector (TOF) is a large cylindrical barrel of Multigap Resistive Plate Chambers (MRPCs) with about 150 000 readout channels, giving very high precision timing for tracks traversing it. Its pseudorapidity coverage matches that of the TPC. Used in combination with the tracking system, the TOF detector can be used for charged particle identification up to about 2.5 GeV/$c$ (pions and kaons) and 4 GeV/$c$ (protons). Still further out from the interaction region, the Electromagnetic Calorimeter (EMCAL) is a Pb--scintillator sampling calorimeter at a distance of $\approx 4.5$ metres from the beam line, 
covering an opening acceptance in the range $|\eta| \leq 0.7$ and $\Delta\phi = 100^\circ$ in azimuth. It has 20.1 radiation lengths and consists of 
11 520 towers.
 
The analysis presented below also makes use of two forward detectors.
 The VZERO counters consist of 
two arrays of 32 scintillator tiles each, covering the range 2.8~$<$ $\eta$ $<$~5.1
(VZERO-A, on the opposite side of the muon arm) and --3.7~$<$~$\eta$~$<$~--1.7
(VZERO-C, on the same side as the muon arm) and positioned respectively at $z$~= 340~cm and $z$~= --90~cm
from the interaction point. 
The Forward Multiplicity Detector~(FMD) consists of Si-strip sensors with a total of 51 240 active detection elements, arranged in five 
rings perpendicular to the beam direction, covering the pseudorapidity ranges $-3.4 < \eta < -1.7$~(FMD-3) and $1.7 < \eta  < 5.1$ (\hbox{FMD-1} and FMD-2), 
a similar coverage to that of the VZERO detector. Finally, 
two sets of hadronic Zero Degree Calorimeters (ZDC) are located at 116 m on either side 
of the interaction point. The ZDCs detect neutrons emitted in the very forward region($|\eta|>8.7$), such as neutrons produced by electromagnetic dissociation~\cite{ALICEEMD}~(see Section 3).
\section{Data analysis}
\subsection{Event selection}
The present analysis is based on a sample of events collected during the
2011 \pb~data-taking, selected with a dedicated barrel ultra-peripheral collision trigger (BUPC), set up to select 
events containing two tracks in an otherwise empty detector. Events 
from two-photon production ($\gamma \gamma \rightarrow \mu^{+}\mu^{-}, e^{+}e^{-}$) or from 
photonuclear vector meson production are selected by this trigger with the 
following characteristics: \newline
(i) at least two hits in the SPD detector; \\
(ii) a number of fired pad-OR ($N^{on}$) in the TOF detector~\cite{TOFtrigger} in the range 2~$\leq N^{on}$$\leq$~6 , 
     with at least two of them with a difference in azimuth,  $\Delta \phi$, in the range $150^\circ \leq \Delta \phi \leq $180$^\circ$; \\
(iii) no hits in the VZERO-A and no hits in the~\hbox{VZERO-C} detectors. \\
A total of about 6.5 $\times 10^{6}$ events were selected by the BUPC trigger.

The integrated luminosity was measured using a trigger for the most central hadronic Pb-Pb collisions.
The cross section for this process was obtained with a van der Meer
scan\ \cite{vanDerMeer}, giving a cross section $\sigma$ = 4.10~$^{+0.22}_{-0.13}$(sys) b~\cite{Kenproceedings}. This gives
an integrated luminosity for the BUPC trigger sample, corrected for trigger live time,
of $\cal L_{\rm int}$ = 23.0~$^{+0.7}_{-1.2}~\mu\rm{b}^{-1}$.
An alternative method based on using neutrons detected in the two ZDCs was also used.
The ZDC trigger condition required a signal in at least one of the two calorimeters, thus selecting single
electromagnetic dissociation (EMD) as well as hadronic interactions.
The cross section for this trigger was also measured with a van der Meer scan, giving a cross section \newline $\sigma$ = 371.4~$\pm$ 0.6(sta)~$\pm ^{24}_{19}$(sys) b\ \cite{ALICEEMD}.
The integrated luminosity obtained for the BUPC by this method is 
consistent with the one quoted above within 3$\%$.

The following selection criteria were applied in the data analysis: \newline
(i) a number of reconstructed tracks 1~$\leq N_{TRK}\leq$~10, where a track is defined with loose criteria:
more than 50$\%$ of findable clusters in the TPC fiducial volume and at least 20 TPC clusters, matching with those found in the ITS;\\
(ii) a reconstructed primary vertex;\\
(iii) only two good tracks passing tighter quality cuts: at least 70 TPC clusters,  
at least 1 SPD cluster and rejection of tracks with a kink.
Moreover the tracks extrapolated to the reconstructed
vertex should have a distance of closest approach~($DCA$) in the longitudinal beam direction $DCA_{L}\leq$~2~cm, and 
$DCA_{T}\leq$0.0182
+0.0350/$p_{\rm T}^{1.01}$ cm in the plane orthogonal 
to the beam direction, where $p_{\rm T}$ is in~(GeV/$c$);\\
(iv) at least one of the two good tracks selected in (iii) with \pt~$\geq$ 1 GeV/$c$; this cut reduces the background while 
it does not affect the genuine leptons from \jpsi~decay;\\ 
(v) The VZERO trigger required no signal within a time window of 25~ns around the collision time 
in any of the scintillator tiles of both VZERO-A and VZERO-C. The time width of the trigger 
windows are limited by the design of the VZERO front-end electronics which is operated at 
the frequency of the LHC clock, $i.e.$~40~MHz.
In the offline analysis the event selection criteria consisted in an absence of a reconstructed 
signal in any of the VZERO scintillator tiles. The time windows in the offline analysis are enlarged to 40~ns and 60~ns around 
the collision time in VZERO-A and VZERO-C, respectively, and were chosen in order to maximize the vetoing efficiency;\\
(vi) the d$E$/d$x$ for the two tracks is compatible with that of electrons or muons; 
Figure 1 shows 
the TPC d$E$/d$x$ of the positive lepton candidate as a function of the d$E$/d$x$ of the negative lepton candidate, 
for \jpsi~candidates in the invariant mass range 2.8~$<$\minv~$<$~3.2 GeV/$c^{2}$. It is worth noting that the TPC 
resolution does not allow to 
distinguish between muons and charged pions;\\
(vii) the two tracks have same or opposite charges, depending on the analysis; \\
(viii) invariant mass 2.2~$<$ \minv~$<$~6~GeV/$c^2$.

The analysis of the $\gamma\gamma$ events is discussed in Section 5. In the remaining of this section
we will focus on \jpsi~analysis.
The effect of the cuts on the statistics is listed in Table~\ref{tab:select}.
In addition to the requirements (i) to (viii),
a first sample enriched with 
coherent events was selected by applying a cut \pt~$<200$~MeV/$c$ for di-muons~(\pt~$<300$~MeV/$c$ for di-electrons).
Photoproduction of vector mesons can occur in interactions where additional photons are exchanged~\cite{neutrons}. 
These additional photons can lead to break up of one or both nuclei. Since the energies of these photons 
are low, only a few neutrons are emitted when the nuclei break up. The exact upper limit on the number of 
emitted neutrons is not known, but in this analysis a cut on the neutron ZDC signal corresponding to less 
than 6 neutrons on each side has been applied. This
cut reduces the statistics by 2.5$\%$, which is considered as a source of systematic error $^{+2.5\%}_{-0\%}$.
After applying all of these selections, 746 di-electron and 1301 di-muon coherent lepton-pair candidates remain.
A second sample was enriched with incoherent events by applying a cut 
\pt~$>200$~MeV/$c$ on di-muons~(\pt~$>300$~MeV/$c$ on di-electrons),
giving 278 electron and 1748 muon incoherent event candidates. 

As described in reference~\cite{aliceupcmuonarm}, during the 2011 Pb-Pb run the VZERO detector was optimized for the
selection of hadronic Pb-Pb collisions, with a threshold corresponding to an
energy deposit above that from a single minimum ionizing particle. Since the VZERO was used 
as a veto in the BUPC trigger, this setting could lead to an
inefficiency in background rejection. In about 30$\%$ of the 2011 BUPC data taking sample, the FMDs were read out too. 
Since these detectors cover a pseudorapidity interval similar to that of the VZEROs, we have used, offline, their information to check for possible inefficiencies
in the VZERO data.
As expected, we found no hits in the FMD detector for the selected BUPC events, confirming that the VZERO inefficiency is very small.

A test on the electron and muon separation was applied to those tracks crossing the EMCAL.
For each track we evaluated the ratio $E/p$ between the energy released in the EMCAL and the reconstructed momentum; electrons
lose their total energy in the shower generated in the EMCAL and for these a value $E/p$~$\approx$~1 is measured.
Minimum ionizing particles lose only a small fraction of their energy in the EMCAL; in this case the measured $E/p$ peaks in the region 0.1-0.2, in good
agreement with the expectation. 
The $e$/$\mu$ separation was obtained by using two methods: (a) a sharp cut on Fig. 1, where all the particles beyond the dotted line 
are considered as electrons, and (b) using the average of the electron~(muon) d$E$/d$x$ and considering as electrons~(muons) the particles within 3 sigma. The difference between
the two methods was used as an estimate of the systematic error~(see Table 2).

\begin{table}
\begin{center}
\begin{tabular}{lr}
Selection                      & Number of \\ 
             & remaining events\\
\hline	
Triggered events                    & 6,507,692 \\
1$\leq N_{TRK} \leq$10                & 2,311,056\\
Primary vertex                                              & 1,972,231 \\
Two reconstructed tracks                                    & 436,720 \\
max($p_{\rm T}^{1},p_{\rm T}^{2})>$1 GeV/$\it{c}$   & 46,324 \\
VZERO offline                                                  & 46,183\\
d$E$/d$x$ consistent with electron~(muon)            & 45,518\\
Opposite sign tracks                                        & 31,529\\
2.2 $<$ \minv~$<$ 6.0~GeV/$c^2$                             & 4,542\\                
\noalign{\smallskip}\hline
\end{tabular}
\end{center}
\caption{\label{tab:select} Summary of the applied data cuts (see text).}	
\end{table}

\subsection{Acceptance and efficiency correction}
The acceptance and efficiency of \jpsi~reconstruction were calculated using a large sample of
coherent and incoherent \jpsi~events generated by STARLIGHT\ \cite{starlightMC} and
folded with the detector Monte Carlo simulation. STARLIGHT simulates 
photonuclear and two-photon interactions at hadron colliders. The simulations
for exclusive vector meson production and two-photon interactions are based on the models
in~\cite{starlight} and~\cite{Baltz:2009jk}, respectively. 

A separate simulation was performed for each run, in order to take into account the slight
variations in run conditions during the data taking.
The product of the acceptance and efficiency corrections $\acceps$~was calculated as
the ratio of the number of the simulated events that satisfy all selections in Table~1 to the
number of generated events with the \jpsi~in the rapidity interval --0.9~$<$y$<$~0.9.
In addition, the reconstructed tranverse momentum is required to be  \pt $<200$~MeV/$c$ (\pt $>200$~MeV/$c$) for
di-muons and \pt $<300$~MeV/$c$ (\pt $>300$~MeV/$c$) for di-electrons in the coherent (incoherent) sample.

\begin{table*}
\begin{center}
\begin{tabular}{lccccc}
Source                                                & Coherent   	 & Incoherent &$\gamma\gamma$~(low) & $\gamma\gamma$~(high)\\ \hline	\\
Luminosity                                            & $^{+5\%}_{-3\%}$ & $^{+5\%}_{-3\%}$ &$^{+5\%}_{-3\%}$  & $^{+5\%}_{-3\%} $   \\
Trigger dead time                                     & $\pm 2.5\%$ & $\pm 2.5\%$& $\pm 2.5\%$ &$\pm 2.5\%$\\
Signal extraction                                     & $^{+7\%}_{-6\%}$\big($^{+6\%}_{-5\%}\big)$ & $^{+26.5\%}_{-12.5\%}$\big($^{+9\%}_{-8\%}$\big)& $\pm 1\%$ & $\pm 4\%$ \\
Trigger efficiency           		                      &  $^{+3.8\%}_{-9.0\%}$&$^{+3.8\%}_{-9.0\%}$&$^{+3.8\%}_{-9.0\%}$&$^{+3.8\%}_{-9.0\%}$	          \\ 		
($\mathrm{ Acc}\times\varepsilon$)               		                  & $\pm 2.5\%$ ($\pm 1\%$)       &  $\pm 6.5(\pm 3.5)\%$  &$\pm 0.3\%$&$\pm 0.5\%$ \\
$\gamma\gamma\rightarrow e^+e^-$ background 		                    & $^{+4\%}_{-0\%}$          &$^{+4\%}_{-0\%}$ & $^{+4\%}_{-0\%}$& $^{+4\%}_{-0\%}$   \\
e/$\mu$ separation                                    &  $\pm 2\%$ & $\pm 2\%$&$\pm 1.7\%$&$\pm 4\%$\\
Branching ratio 				                              &  $\pm 1\%$            & $\pm 1\%$&-&-\\ 
Neutron number cut			                                              &  $^{+2.5\%}_{-0\%}$            & - &-&-\\  
Hadronic \jpsi                                        &   -& $^{+0\%}_{-5\%}$ \big($^{+0\%}_{-3\%}$\big) &-& - &\\
\hline \\
Total                                                 &  $^{+14.0\%}_{-9.6\%}$\big($^{+13.4\%}_{-8.8\%}$\big) & $^{+29.4\%}_{-16.6\%}$\big($^{+14.5\%}_{-11.7\%}$\big)
&  $^{+10.8\%}_{-7.0\%}$ & $^{+12.0\%}_{-8.8\%}$
\\ 
\end{tabular}
\end{center}
\caption{\label{tab:sss} Summary of the contributions to the systematic error for
the \jpsi~and $\gamma\gamma$~cross section measurement for electrons~(muons). The error for the \jpsi~signal extraction includes
the systematic error in the fit of the invariant mass spectrum and the systematic errors on
$f_D$ and  $f_I$($f_C$), as described in the text.}	
\end{table*}

\begin{table*}
\begin{center}
\begin{tabular}{lcccccc}
Sample       & Coherent   	                    & Coherent                    & Incoherent                        & Incoherent   & $\gamma\gamma\rightarrow e^{+}e^{-}$ & $\gamma\gamma\rightarrow e^{+}e^{-}$\\ 
             & enriched                         & enriched                    & enriched                          & enriched                    &                                      &                    \\  
             &\jpsi$\rightarrow \mu^{+}\mu^{-}$ &\jpsi$\rightarrow e^{+}e^{-}$& \jpsi$\rightarrow \mu^{+}\mu^{-}$ &\jpsi$\rightarrow e^{+}e^{-}$& (low)&(high) \\
\hline
Yield                                           &$291\pm 18$(sta)& $265\pm 40$(sta)& $91\pm 15$(sta)&$61\pm 14$(sta) & $186\pm 13$(sta)& $93\pm 10$(sta)  \\
                                   &$\pm$ 4(sys) &$\pm$ 12(sys) & $^{+7}_{-5}$(sys) &$^{+16}_{-7}$(sys) &$\pm$ 12(sys) &$\pm$ 6(sys)   \\
Mass(GeV/$c^{2}$) &3.096$\pm$0.002 & 3.092$\pm$0.036&3.085$\pm$0.007 & 3.080$\pm$0.007&-&-\\
$\sigma$ (MeV/$c^{2}$)   &25$\pm$ 1.1&25.0$\pm$ 1.9&33$\pm$ 6&25.0$\pm$ 1.4&-&-   \\
$\mathrm{ Acc}\times\varepsilon$ (\%)           		                     	&4.57&2.71&3.19&1.8&5.6&4.73   \\
LS pairs               		                 &3&0&53&8&0&0   \\
OS pairs 		                   &365&514&178&143&186&93   \\
$f_{D}$                                   &$0.1^{+0.05}_{+\-0.06}$&$0.1^{+0.05}_{+\-0.06}$&0.095$\pm$0.055&0.11$\pm$0.07&- &-   \\ 
$f_{I}$			                               &0.044$\pm$0.014&0.15$\pm$0.02&-&-&-&-   \\
$f_{c}$			                               &-&-&0.03$\pm$0.03&0.47$\pm$0.09&-&-   \\     
$\lambda_{\gamma\gamma}$(GeV$^{-1}~c^2)$  &-&-&-&-& -1.55$\pm$0.88  & -0.73$\pm$0.18  \\
\hline
\\ 
\end{tabular}
\end{center}
\caption{\label{tab:syserr} Summary of the main experimental results obtained in the \jpsi~and $\gamma\gamma$~analysis and of the most relevant correction 
parameters applied.}	
\end{table*}

The average values for the combined acceptance and efficiency for $\jpsi \rightarrow e^{+}e^{-}(\mu^{+}\mu^{-}$)
were found to be 2.71\ (4.57)$\%$ and 1.8\ (3.19)$\%$ for 
coherent and incoherent \jpsi, respectively. 

The STARLIGHT model predicts a dependence of the \jpsi~cross section on the rapidity, giving 
a $\approx$10$\%$ variation over the rapidity range $y$=$\pm$0.9.
In order to evaluate the systematic error on the acceptance coming from the generator choice,
we used a flat dependence of $\mathrm{d}\sigma_{J/\psi}/\mathrm{d}y$ in the interval 
-0.9~$<$y$<$0.9, as predicted by other models (see Fig. 5).
The relative differences in ($\mathrm{ Acc}\times\varepsilon$) between the methods were 2.5\ (1.0)\% for coherent electrons~(muons), and 
6.5\ (3.5)\% for incoherent electrons~(muons), and are
taken into account in the systematic error calculation.
Transverse polarization is expected from helicity conservation for a quasi-real photon.
It is therefore assumed in these calculations that the \jpsi~is transversely
polarized, as found by previous experiments~\cite{Chekanov:2002xi,Aktas:2005xu}.
The trigger efficiency was measured relying on a data sample collected in a dedicated run
triggered by the ZDCs only. 
We selected events with a topology having the BUPC conditions, given at the beginning of section 3.1.
The resulting trigger efficiency was compared with that obtained by the
Monte Carlo simulation, showing an agreement within $^{+3.8\%}_{-9.0\%}$.
\subsection{Analysis of invariant mass spectrum}
Figure 2 shows
the invariant mass distribution for 2.2 $<$ \minv~$<$ 6.0 GeV/$c^{2}$
for opposite-sign (OS) and like-sign (LS) electron and muon pairs.
A \jpsi~peak is clearly
visible in the four spectra, on top of a continuum coming from
$\gamma \gamma \rightarrow e^{+}e^{-}(\mu^{+}\mu^{-})$ for the coherent enriched sample.
The continuum for the incoherent enriched sample for the muon channel (bottom, left) is likely 
to come from misidentified  $\pi^+ \pi^-$ pairs.
To extract the \jpsi~yield,
the number 
of OS events in the interval 2.2 $<$\minv~$<$ 3.2 GeV/$c^{2}$ for electrons and 
3.0 $<$ \minv~$<$ 3.2 GeV/$c^{2}$ for muons were considered.
In the mass intervals quoted above, 0\ (3) LS electron(muon)-pairs were found
for coherent enriched events, 
while 8\ (53)  
LS pairs were found for incoherent enriched events.
The corresponding number of
OS pairs was 514\ (365) for coherent enriched sample and 143\ (178) for incoherent enriched events.
The \jpsi~yield was obtained by fitting the di-lepton invariant mass spectrum 
with an exponential function
to describe the underlying continuum, and a Crystal Ball function\ \cite{Gai:CrystalBall}
to extract the \jpsi~signal. The Crystal Ball function tail parameters ($\alpha_{CB}$ and $n$) were left free for the
coherent enriched sample, giving a good agreement with those obtained by fitting the simulated data, and were fixed to values
obtained from simulations for the incoherent enriched one.
The background found in the incoherent sample was taken into account into the fit by using a $5^{th}$ order polynomial in addition to the Crystal Ball 
and an exponental function. This contribution was normalized according to the experimental LS pair spectrum~(Fig. 2).

\section{The \jpsi~cross section}
\subsection{Coherent \jpsi~cross section}
The yield obtained for the coherent enriched sample~(Fig. 2~top) is
$N_{\rm yield} = 265 \pm 40\rm{(sta)}\pm 12\rm{(sys)}$ for the $\jpsi \rightarrow e^{+}e^{-}$ channel and
$N_{\rm yield} = 291 \pm 18\rm{(sta)}\pm 4\rm{(sys)}$ for the $\jpsi \rightarrow \mu^{+}\mu^{-}$ channel.
The systematic error on the yield 
is obtained by varying the bin size and by replacing the exponential with a polynomial to fit the $\gamma\gamma$ process.
In addition, the Crystal Ball function parameters are obtained by fitting a simulated sample made of \jpsi~and~$\gamma\gamma$ event cocktail 
and then used to fit the coherent enriched data sample too. The difference in the yield obtained with the two Crystal Ball fit 
procedures is included in the systematic error.
As a result we obtain a $^{+7\%}_{-6\%}$ and $^{+6\%}_{-5\%}$  systematic error on the signal extraction for coherent electrons and muons, respectively.
For the coherent enriched sample,  
the central mass (width calculated from the standard deviation) value from the fit is 
$3.092 \pm 0.036$~GeV/$c^2$\ (25.0$\pm$1.9 MeV/$c^{2}$)~for the electron channel and 
$3.096 \pm 0.002$~GeV/$c^2$\ (25$\pm$1.1 MeV/$c^{2}$) for the muon channel,
in good agreement with 
the known value of the \jpsi~mass and compatible with the absolute calibration 
accuracy of the barrel.
For the incoherent sample,  
the central mass~(width calculated from the standard deviation) value from the fit is 
$3.080 \pm 0.007$~GeV/$c^2$\ (25$\pm$1.4 MeV/$c^{2}$) for the electron channel and
$3.085 \pm 0.007$~GeV/$c^2$\ (33$\pm$6 MeV/$c^{2}$) for the muon channel.
The exponential slope parameter, $\lambda_{\gamma\gamma}$, of the
continuum for the coherent enriched sample is computed at \mbox{2.2 $<$ \minv~$<$ 2.6 GeV/$c^{2}$~(low)} and 
3.7 $<$ \minv~$<$ 10 GeV/$c^{2}$~(high) for electrons 
with -0.9$<$$\eta_{1,2}$$<$0.9, giving -1.55$\pm$0.88~GeV$^{-1}~c^2$
and \newline -0.73$\pm$0.18~GeV$^{-1}~c^2$, in 
good agreement with the corresponding Monte Carlo
expectation,\newline -1.07$\pm$0.16~GeV$^{-1}~c^2$ 
and -0.81$\pm$0.01~GeV$^{-1}~c^2$, respectively.
This is an additional indication that there is no important
background in the invariant mass and \pt~range considered. 

Exclusive photoproduction of $\psi^{'}$, followed by the $\psi^{'}\rightarrow \jpsi$~+~anything 
decay, can be a background for this analysis 
when particles produced in addition to the \jpsi~are undetected.
The fraction $f_{D}$ of coherent \jpsi~mesons coming from the decay 
$\psi^{'}\rightarrow \jpsi$~+~anything, was estimated 
following the same prescription used in~\cite{aliceupcmuonarm},
with the 
theoretical estimates for $f_D$ 
ranging from 4.4$\%$ to 11.8$\%$ for electrons and
4.3$\%$ to 14.7$\%$ for muons. 
Alternatively, the ratio of coherent yields for $\psi^{'}$ to $J/\psi$  
can be extracted from the real data. Owing to the limited statistics, 
we combine the  electron and muon channels to
obtain  $N_{\psi^{\prime}}$=17$\pm$10 and $N_{\psi}$=505$\pm$48~(see Fig. 3).
The fraction $f_D$, for a given \jpsi~polarization $P$, can be written as:
\begin{equation}
\label{eq2a2}
\begin{split}
f_{D}^{P}
=\frac
{
N_{\psi^{'}}\cdot
(\mathrm{ Acc}\times\varepsilon)_{\psi^{'}\rightarrow\jpsi}^{P}
}
{
(\mathrm{Acc}\times\varepsilon)_{\psi^{'}\rightarrow \l^{+}l^{-}}
}
%
{~~~~~~~~~~~~~~~~~~~~~~~~~~~~~~~~~~~~~~~~~~~~}
\\
\times
\frac{
BR(\psi^{'} \rightarrow \jpsi+\mathrm{anything})
}{
BR(\psi^{'} \rightarrow l^{+}l^{-})
}
{~~~~~~~~~~~~~~~~~~~~~~~~~~~~~~~~~~~~~~}
\\
\times
\frac
{BR(\jpsi \rightarrow l^{+}l^{-})}
{N_{\jpsi}},
{~~~~~~~~~~~~~~~~~~~~~~~~~~~~~~~~~~~~~~~~~~~~~~~~~~}
\end{split}
\end{equation}
where  $(\mathrm{ Acc}\times\varepsilon)_{\psi^{'}\rightarrow\jpsi}^{P}$ ranges from
2 to 3$\%$ for electrons and from 3.4 to 4.6 $\%$ for muons, depending on the \jpsi~polarization.
The ($\mathrm{Acc}\times\varepsilon)_{\psi^{'}\rightarrow \l^{+}l^{-}}^{P}$ ranges from
3.3 to 4.5$\%$ for electrons and muons, respectively.
The acceptance corrections are polarization dependent and 
give $f_{D}^{P}$ ranging from 15$\pm$9$\%$ for longitudinal polarization to 
11$\pm$6.5$\%$ for transverse polarization.

In what follows, we use the central value of theoretical and experimental estimates, and take the
others as upper and lower limits, \textit{i.e.}~$f_{D}$=$0.10^{+0.05}_{-0.06}$.
The di-electron~(di-muon) \pt~distribution, integrated over 2.2~$<$\minv~$<$~3.2~GeV/$c^2$,
~(3.0~$<$\minv~$<$~3.2~GeV/$c^2$)
is shown in Fig.~4 right~(left). 
The clear peak at
low \pt~is mainly due to coherent interactions, while the tail extending 
out to 1 GeV/$c$
comes from incoherent production. 
To estimate the fraction ($f_I$) of incoherent over coherent events in 
the low~\pt\\ region 
(\pt $<300$~MeV/$c$ for di-electrons,~\pt $<200$~MeV/$c$ for di-muons), the 
ratio $\sigma_\mathrm{inc}/\sigma_\mathrm{coh}$, weighted by the detector
acceptance and efficiency for the two processes, was calculated, giving 
$f_{I}$=0.13~(0.06) for di-electrons~(di-muons) when
$\sigma_\mathrm{inc}/\sigma_\mathrm{coh}$ was taken from STARLIGHT, and 
$f_{I} =0.05~(0.03)$ when the
model in~\cite{Rebyakova:2011vf} was used with leading twist contribution.
For higher twist contributions the above model gives $f_{I} =0.07~(0.03)$. 
An alternative method to extract an upper limit of $f_I$ from the data was carried out by
fitting the measured \pt~distribution.
Six different functions were used to describe the \pt~spectrum:\\
(i)~coherent \jpsi~photoproduction; \\
(ii)~incoherent \jpsi~photoproduction;\\
(iii)~\jpsi~from coherent $\psi^{'}$ decay;\\
(iv)~\jpsi~from incoherent $\psi^{'}$ decay;\\
(v)~two-photon production of continuum pairs;\\
(vi)~\jpsi~produced in peripheral hadronic collisions.

 The shapes for the first five fitting functions (Monte Carlo templates) were provided by
STARLIGHT events folded with the detector simulation, while the last one is extracted from data
at higher centralities\cite{centrality}. The relative normalization was left
free for coherent and incoherent photoproduction.
The contribution from the $\psi^{'}$ was constrained from the estimate above ($f_{D}$=$0.10^{+0.05}_{-0.06}$), 
and the two-photon contribution was determined from the fit to the
continuum in Fig.~2. 
The hadronic \jpsi~were constrained by the fit to the region $p_{T}$$>$1.1~GeV/$c$, where the ultra-peripheral \jpsi~contribution is negligible. 
As a result of the fit, we obtain  $f_{I}$=$0.044\pm 0.014$ for di-muons and
$f_{I}$=$0.15\pm 0.02$ for di-electrons. 
Since these values are compatible, within the errors, with the theoretical expectations 
(for both models in the case of di-muons and for STARLIGHT for the di-electrons), they are used in the calculations.
The fit reproduces properly the experimental data $p_{T}$ spectrum, clarifying the origin of the 
high $p_{T}$ {\jpsi}s~pointed out in the PHENIX paper~\cite{Afanasiev:2009hy}.
Finally, the total number of coherent \jpsi s is calculated from the yield extracted from the fit to the
invariant mass distribution by
\begin{equation}
N^{\mathrm{coh}}_{\jpsi} = \frac{N_{\rm yield}}{1 + f_I + f_D} \; ,
\label{NCohJPsi}
\end{equation}
resulting in $N^{\mathrm{coh}}_{\jpsi}(\mu^{+}\mu^{-})  = 255\pm 16(\mathrm{sta}) ^{+14}_{-13} (\mathrm{sys}) $ and
$N^{\mathrm{coh}}_{\jpsi}(e^{+}e^{-}$)  =$212 \pm 32(\mathrm{sta}) ^{+14}_{-13} (\mathrm{sys}) $
respectively.
The coherent \jpsi~differential cross section is given by:
\begin{equation}
\frac{\mathrm{d}\sigma_{\jpsi}^{\mathrm{coh}}}{\mathrm{d}y}  =
\frac{N_{\jpsi}^{\mathrm{coh}}}
{\acceps 
\cdot BR(\jpsi \rightarrow l^{+}l^{-}) \cdot {\cal{L}_{\mathrm{int}}} \cdot\Delta y  } \; ,
\label{eq1a}
\end{equation}
where $N^{\mathrm{coh}}_{\jpsi}$ is the number of \jpsi~candidates from Eq.~\ref{NCohJPsi} and
$\acceps$~corresponds to the acceptance and efficiency as discussed above.
$BR(\jpsi \rightarrow l^{+}l^{-})$ is the branching ratio for \jpsi~decay into leptons\ \cite{Beringer:1900zz},
$\Delta \it{y} =$~1.8 the rapidity interval bin size, and $\cal{L}_{\mathrm{int}}$
the total integrated luminosity.
As a result we obtain
$\mathrm{d}\sigma_{J/\psi}^{\mathrm{coh}} /\mathrm{d}y = 2.27 \pm 0.14(\mathrm{sta})^{+0.30}_{-0.20}(\mathrm{sys})$~mb for the di-muon channel and
$\mathrm{d}\sigma_{J/\psi}^{\mathrm{coh}} /\mathrm{d}y = 3.19\pm 0.50(\mathrm{sta})^{+0.45}_{-0.31}(\mathrm{sys})$~mb for the electron one.
Since the di-electron and di-muon data are statistically separated samples, they can be combined; their 
weighted average gives 
$\mathrm{d}\sigma_{J/\psi}^{\mathrm{coh}} /\mathrm{d}y =\\ 
2.38^{+0.34}_{-0.24}\big(\mathrm{sta+sys}\big)$~mb.

In addition, the fraction $F_{n}$ of coherent events with no neutron emission was estimated by STARLIGHT to be 
$F_{n}$=0.68, while the model~\cite{Rebyakova:2011vf} predicts $F_{n}$=0.76. Events 
with neutron emission can be efficiently tagged in ALICE by the ZDC calorimeters, 
taking advantage of their high efficiency ($>$98$\%$).
By fitting the di-electron (di-muon) invariant mass spectrum for events with and without neutron emission and with
\pt $<300$~MeV/$c$\ (\pt $<200$~MeV/$c$), we obtain a fraction $0.70\pm 0.05\rm{(sta)}$ in good agreement 
with the above estimates.

\subsection{Incoherent \jpsi~cross section}
The incoherent cross sections are obtained in a similar way. 
For the incoherent enriched sample the obtained yield is~(Fig.2~bottom), 
$N_{\rm yield} =61 \pm 14\rm{(sta)}^{+16}_{-7}\rm{(sys)}$ for the $\jpsi \rightarrow e^{+}e^{-}$ channel and 
$N_{\rm yield} = 91 \pm 15\rm{(sta)}^{+7}_{-5}\rm{(sys)}$ for the 
$\jpsi \rightarrow \mu^{+}\mu^{-}$ channel.
Here $f_{D}$ represents the fraction of incoherent \jpsi~mesons coming from the decay 
$\psi^{'}\rightarrow \jpsi$~+~anything, 
and was obtained only from formula (1), since
the limited statistics did not allow the extraction of the $\psi^{'}$ yield from the data. 
The predictions for incoherent $f_{D}$ are calculated using both STARLIGHT and the model~\cite{Rebyakova:2011vf}.
As a result we obtain a $f_{D}$ value ranging from 3.9$\%$ to 15.1$\%$ for muons and
from 3.8$\%$ to 18.1$\%$ for electrons. By using the average we obtain
$f_{D}=(9.5\pm 5.5)\%$ for muons and $f_{D}=(11\pm 7)\%$ for electrons. 
Using STARLIGHT, $f_{C}$~(the fraction of coherent \jpsi~contaminating the incoherent sample), corrected by 
the acceptance and the efficiency, is found to be $f_{C}$=0.5 for electrons and $f_{C}$=0.02 for muons. 
By fitting the measured \pt~distribution (Fig. 4) we extract $f_{C}$=$(0.47\pm 0.09)$ 
for electrons, while $f_{C}$ is $f_{C}$=$(0.03\pm 0.03)$ for muons. These results are compatible with 
those from the models and will be used in the following.
By applying the ratio 1/(1+$f_{D}$+$f_{C}$) to the $N_{\rm yield}$, 
the total number of incoherent muon events is    
$N^{\mathrm{inc}}_{\jpsi}(\mu^{+}\mu^{-})~=81\pm 13(\mathrm{sta})^{+8}_{-6} (\mathrm{sys}) $, corresponding to
$\mathrm{d}\sigma_{J/\psi}^{\mathrm{incoh}} /\mathrm{d}y = 1.03 \pm 0.17(\mathrm{sta})^{+0.15}_{-0.12}(\mathrm{sys})$~mb for the di-muon channel.
For electrons 
we obtain $N^{\mathrm{inc}}_{\jpsi}(e^{+}e^{-}) =   39\pm9(\mathrm{sta})^{+10}_{-5}(\mathrm{sys})$, 
corresponding to 
$\mathrm{d}\sigma_{J/\psi}^{\mathrm{incoh}} /\mathrm{d}y = 0.87\pm 0.20(\mathrm{sta})^{+0.26}_{-0.14}(\mathrm{sys})$~mb for the di-electron channel. 
Since these are statistically separate channels, their weighted average gives 
$\mathrm{d}\sigma_{J/\psi}^{\mathrm{incoh}} /\mathrm{d}y = 
0.98^{+0.19}_{-0.17}\big(\mathrm{sta+sys}\big)$~mb.
\subsection{Background and systematic error estimate}
As discussed in~\cite{aliceupcmuonarm}, a possible loss of events might come from correlated QED pair production, 
\textit{i.e.}
interactions which produce both a \jpsi~and a low mass $e^+ e^-$ pair (the latter process has a
very large cross section), with one of the electrons hitting the VZERO detector and thus
vetoing the event.
This effect was studied in~\cite{aliceupcmuonarm}, with a control data sample where no veto at trigger level
was applied.
As a result, an upper limit on the inefficiency smaller than 2\% was found. 
In the forward rapidity trigger only VZERO-A was used as a veto,
and therefore we estimate, conservatively, a 4\% systematic error for this study.

Another possible source of systematic error is the radiative decay 
\jpsi$\rightarrow e^+ e^-$, 
neglected by the event generator used in this paper.
We simulated a $\jpsi\rightarrow e^+e^-$ sample, where 15\% of the events had a photon in the final state~\cite{erratum}.
The Crystal Ball function fit applied to this sample provides fit parameters identical to those of the standard sample, and 
the $(\mathrm{Acc}\times\varepsilon)$ is also not distinguishable from the standard value, so no correction is required in this analysis.

A possible background from hadronic \jpsi~is found (Fig. 4) to be negligible for   
\pt\ below around 200-300 MeV/$c$, and therefore it is not important for coherent production. 
For incoherent events this background was evaluated from the \pt~fit described above and gives a contribution 
(0.043$\pm$0.015) for di-electrons and\\ (0.024$\pm$0.017) for di-muons. 
These fractions refer to events in the mass interval 2.2 $<$\minv~$<$ 3.2 GeV/$c^{2}$ for di-electrons and
3.0 $<$\minv~$<$ 3.2 GeV/$c^{2}$ for di-muons respectively, 
and therefore are not necessarily \jpsi~only. 
We use these fractions as upper limits to be included in the systematic error, giving a contribution $^{+0\%}_{-5\%}$ and 
$^{+0\%}_{-3\%}$, respectively.
The hadronic combinatorial background can be estimated by LS events~(see Table 2). It is negligible for coherent events and for incoherent di-electrons.
For incoherent di-muons this background, possibly coming from misidentified pion pairs, was taken into account by using a polynomial 
function in the corresponding fit, as described at the end of Section 3.

Another source of background may come from photo-produced \jpsi~by nuclei with impact parameters 
$b$$<$2$R$. According to a simulation, based on a calculation method similar to STARLIGHT, the cross section 
for this process (usually not included in the event generator) is 1.1~mb and  0.7~mb in the centrality bin (80-90)$\%$ and
(90-100)$\%$, respectively. The survival probability of the events in these two bins was  
simulated with $2.2\cdot10^{6}$ Pb-Pb minimum bias events produced by the HIJING event generator.
Assuming the trigger conditions (i,~ii,~Section 3) and the analysis cuts (ii,~iv,~vi, Section 3) to be fully satisfied by di-leptons
produced in UPC \jpsi~decays, we find the fraction of events passing the trigger cut (iii) and the analysis cuts (iii,~v) to be 
0.06$\%$ and 0.3$\%$ in the two centrality bins. This process therefore gives a negligible contribution
to the ultra-peripheral cross section.

\section{Two-photon cross section}
The STAR Collaboration measured the two-photon cross section with a precision of 22.5$\%$
when adding the statistical and systematic errors in quadrature~\cite{Adams:2004rz}. This result was slightly larger 
than the one predicted by STARLIGHT, but within $\sim 2\sigma$.
The PHENIX Collaboration has also measured the cross section of two-photon production 
of di-electron pairs\cite{Afanasiev:2009hy}. This measurement, which has an uncertainty of about $30\%$, 
when the statistical and systematic errors are added in quadrature, was found to be in good agreement
with STARLIGHT.  

The cross section for 
$\gamma \gamma \rightarrow e^+ e^-$ 
can be written in a similar
way to Eq.~3,

\begin{equation}
\sigma_{\gamma\gamma}   =
\frac{N_{\gamma\gamma}}
{
(\mathrm {Acc}\times\varepsilon)_{\gamma\gamma}\cdot 
 \cal L_{\rm int}},
\label{comp2}
\end{equation}
where $N_{\gamma\gamma}$ was obtained by fitting the continuum in the invariant mass intervals 
2.2  $<$ \minv~$<$ 2.6~GeV/$c^{2}$ ($N_{\gamma\gamma}^{e^+ e^-}=~186\pm 13\rm{(sta)}\pm 4\rm{(sys)}$) and
3.7 $<$ \minv~$<$ 10~GeV/$c^{2}$ ($N_{\gamma\gamma}^{e^+ e^-} =~93\pm 10\rm{(sta)}\pm 4\rm{(sys)})$, 
to avoid contamination from the \jpsi~peak. In this analysis the integrated luminosity used 
was $\cal L_{\rm int}$ = 21.7~$^{+0.7}_{-1.1}~\mu\rm{b}^{-1}$ and the cut (iv) on the track \pt~was removed.
The cross section for the process $\gamma \gamma \rightarrow \mu^+ \mu^-$ was not studied 
due to a possible contamination~(although small) from pions in the di-muons sample, suggested by the presence 
of LS events.
The cross section
for di-lepton invariant mass was computed between 2.2  $<$ \minv~$<$ 2.6~GeV/$c^{2}$ and 
3.7 $<$ \minv~$<$ 10~GeV/$c^{2}$, for a 
di-lepton rapidity\ in the interval -0.9 $< \it{y} <$ 0.9, and requiring 
$-0.9<\eta_{1,2}<0.9$ for each lepton.
The data cuts applied to the Monte Carlo sample are the same as those applied in
the analysis described above, resulting in a $(\mathrm{Acc}\times\varepsilon)_{\gamma\gamma}^{e^+ e^-}$=5.6$\%$ 
for 2.2 $<$ \minv~$<$ 2.6~GeV/$c^{2}$ and 
$(\mathrm{Acc}\times\varepsilon)_{\gamma\gamma}^{e^+ e^-}$=4.73$\%$ for 3.7 $<$ \minv~$<$ 10~GeV/$c^{2}$.  
As a result we obtain
$\sigma_{\gamma \gamma}^{e^+ e^-}$ = $154\pm 11(\mathrm{sta})^{+17}_{-11}(\mathrm{sys})$~$\mu$b 
for the lower invariant mass interval and\\
$\sigma_{\gamma \gamma}^{e^+ e^-}$ = $91\pm 10(\mathrm{sta})^{+11}_{-8}(\mathrm{sys})$~$\mu$b
for the higher invariant mass interval, 
to be compared with $\sigma$=128~$\mu$b and $\sigma$=77~$\mu$b~given by STARLIGHT, respectively.
In Fig.~6 the invariant mass distributions for\\ 2.2~$<$ \minv~$<$ 2.6~GeV/$c^{2}$ interval and for  
3.7 $<$ \minv~$<$ 10~GeV/$c^{2}$
are shown.
\section{Discussion}
The cross section of coherent \jpsi~photoproduction is compared with
 calculations from six different models\ \cite{Rebyakova:2011vf,starlight,Adeluyi:2012ph,Goncalves:2011vf,Cisek:2012yt,LM}
in Fig. 5(a). The incoherent production cross section is compared with calculations by three 
different models\ \cite{Rebyakova:2011vf,starlight,LM}.
These models calculate the photon spectrum in impact parameter 
space in order to exclude interactions where the nuclei interact hadronically. The differences between the
models come mainly from the way the photonuclear interaction is treated. The predictions can be divided into
three categories: \newline
i) those that include no nuclear effects (AB-MSTW08, see below for definition). In this
approach, all nucleons contribute to the scattering, and the forward scattering differential cross section,
$\mathrm{d} \sigma/\mathrm{d} t$ at $t = 0$ ($t$ is the momentum transfer from the target nucleus squared),
scales with the number of nucleons squared, $A^2$; \\
ii) models that use a Glauber approach to calculate the number of nucleons
contributing to the scattering (STARLIGHT, GM, CSS and LM). The calculated cross section depends on the 
total \jpsi-nucleon cross section and on the nuclear geometry; \\
iii) partonic models, where the cross section is proportional to the nuclear gluon
distribution squared (AB-EPS08, AB-EPS09, AB-HKN07, and RSZ-LTA).

The rapidity region -0.9~$<$$y$$<$~0.9 considered here corresponds to photon-proton 
centre--of--mass energies, $W_{\gamma\mathrm{p}}$, between 59 and 145 GeV. The corresponding 
range in Bjorken-{\it x} is between $x = 5 \times 10^{-4}$ and $x = 3 \times 10^{-3}$. 
In this region, a rather strong shadowing is expected, and models based on perturbative QCD predict a 
lower value for the cross section than models using a Glauber approach to account for the nuclear effect. 

The measured cross section, $\mathrm{d}\sigma_{J/\psi}^{\mathrm{coh}} 
/\mathrm{d}y =\\
 2.38^{+0.34}_{-0.24}\big(\mathrm{sta+sys}\big)$~mb 
is in very good agreement with the calculation by Adeluyi and Bertulani 
using the EPS09 nuclear gluon prediction. The GM model, and the other models using 
a Glauber approach, predict a cross section a factor 1.5-2 larger than the data, 
overestimating the measured cross section by more than 3 standard 
deviations. So does the prediction based on the HKN07 parametrization, which includes 
less gluon shadowing than EPS09. 

The model AB-EPS08, significantly underestimates the measured
cross section by about a factor of two (about 5 standard deviations), indicating that the gluon shadowing is too strong in the
EPS08 parameterization. The leading twist calculation (RSZ-LTA) is also significantly below the data, by about 2-3 sigma.

For the incoherent cross section, shown in fig. 5(b), there are three model predictions available, LM,\\
STARLIGHT, and RSZ-LTA. The measured value deviates by about two standard deviations from the LM prediction, while STARLIGHT predicts an incoherent cross section 60$\%$ too high, and RSZ-LTA a factor 4 too low. Taking the ratio between the incoherent and coherent cross section provides further constraints on the treatment of the nuclear modifications implemented in the different models. Another advantage is that the photon spectrum is factorized out, so that the comparison directly probes the ratio of the photonuclear cross sections. The ratio obtained from data is\\
$0.41^{+0.10}_{-0.08}\big(\mathrm{sta+sys} \big)$. This can be compared with 0.21 from LM, 0.41 from STARLIGHT,  and 0.17 from RSZ-LTA. Although the RSZ-LTA model is quite close for the coherent cross section at mid-rapidity, it seems to underpredict the incoherent cross section. The LM model also predicts a too low ratio. STARLIGHT, on the other hand, has about the right ratio of incoherent-to-coherent cross section, although it does not reproduce any of the cross sections individually. All three models use the Glauber model to calculate the incoherent cross section, but the implementation and the input cross section for $\gamma+p \rightarrow J/\psi + p$ varies.
In STARLIGHT the scaling of the inelastic \jpsi~+ nucleus cross section, ranges from $A^{2/3}$ to $A$,  depending on the \jpsi~+~nucleon cross section.
In the first case, only the nucleons on the surface participate in the scattering, while in the second one all the nucleons contribute.
The cross section for incoherent photoproduction is assumed in STARLIGHT to follow the same scaling, while in the other models, the reduction with respect to the $A$ scaling is larger.

The measured values for the $\gamma\gamma$ cross sections are 20$\%$ above but fully compatible within 1.0  and 1.5 sigma with the STARLIGHT prediction for the high and low 
invariant mass intervals, respectively, 
if the statistical and systematic errors are added in quadrature. 
This 
result provides important constraints on calculations that include terms of higher orders in $\alpha_{em}$. 
A reduction in the two-photon cross section 
of up to \mbox{30 $\%$} compared with leading-order calculations has been  
predicted~\cite{Baltz:2009fs}. 
The result for the two-photon cross section to di-lepton pairs, measured by ALICE with a
 precision of 12$\%$ and 16$\%$ for the low and high invariant mass range respectively,  is thus fully consistent 
with STARLIGHT, and sets limits on the contribution from higher order terms~\cite{Hencken:2006ir}.
This result supports the ALICE \jpsi~photoproduction measurement in the forward rapidity region~\cite{aliceupcmuonarm}, where the cross 
section was based on $\sigma_{\gamma \gamma}$. 

\section{Summary}
In summary, the first measurement of coherent and incoherent \jpsi~photoproduction and 
two-photon production of di-lepton pairs in Pb-Pb collisions at mid-rapidity at the LHC has been presented and compared with model 
calculations. The \jpsi~photoproduction cross sections provide a powerful tool to constrain the nuclear gluon shadowing in the region 
$x$~$\approx 10^{-3}$. The coherent \jpsi~cross section is found to be in good agreement with the model 
which incorporates the nuclear gluon shadowing according to the EPS09 parameterization (AB-EPS09). 

Models which include no nuclear gluon shadowing are inconsistent with the measured results, as those which use the 
Glauber model to incorporate nuclear effects. The AB-HKN07 and AB-EPS08 models contain too little or 
too much shadowing, respectively, to match the data.
Our results are about 3 sigma higher than the RSZ-LTA model prediction, although a deviation of just 1.5 sigma is found
from the model upper limit.
Nevertheless the above predictions may have large uncertainties coming not only from the  parametrization 
of the nuclear gluon distribution but also from the selection of the hard scale, the contributions 
from the higher order terms and the treatment of the photon fluctuation to a quark-antiquark pair. 
The current measurement will contribute to resolve these uncertainties.

None of the three existing models predicts the incoherent photoproduction cross section correctly, 
but STARLIGHT predicts a correct incoherent-to-coherent ratio. 
 
Finally, the measured two-photon cross section for di-electron production 
is consistent with the\\ STARLIGHT model. This implies the models predicting a strong 
contribution of higher-order terms (not included in STARLIGHT) to the cross section are not favored. 

\begin{figure}[htbp]
\centering
{\includegraphics[width=0.8\linewidth,keepaspectratio]{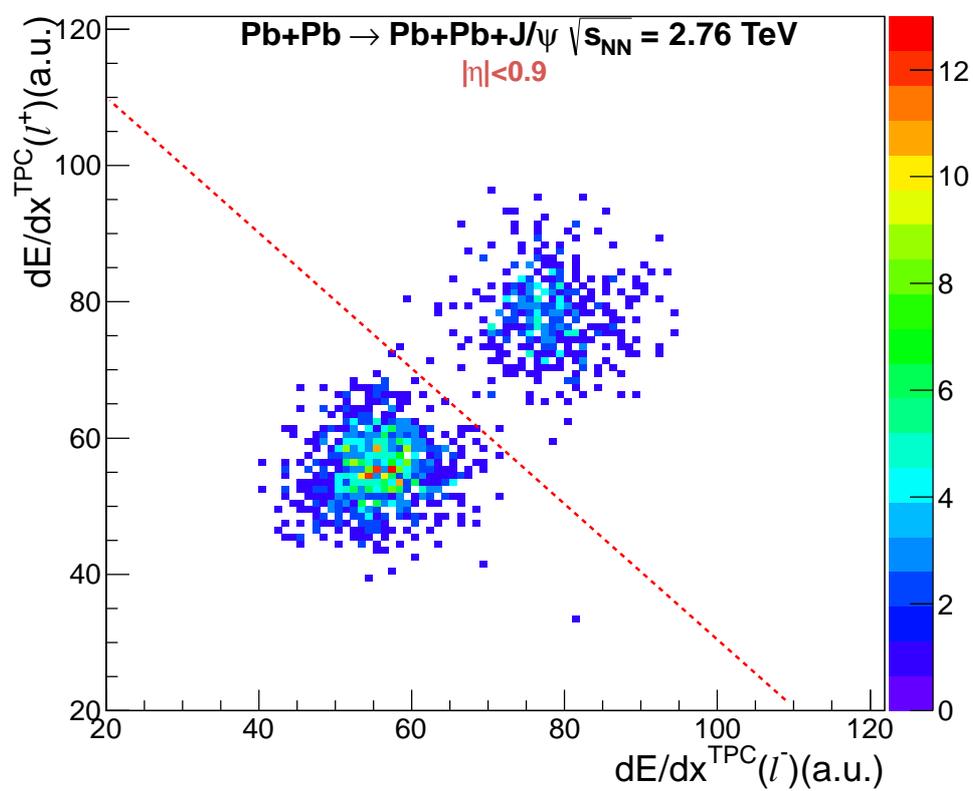}}
\caption{\label{fig:dedx}d$E$/d$x$ of the positive lepton versus the negative one, 
as measured by the TPC 
for \jpsi~candidates in the ultra-peripheral Pb-Pb collisions at $\sqrt{s_{\mathrm{NN}} } = 2.76$ TeV in the invariant mass range 2.8 $<$~\minv~$<$ 3.2 GeV/$c^{2}$ and -0.9~$<\eta<$~0.9.
Muon pairs and electron pairs are clearly separated, 
with the latter showing higher d$E$/d$x$ values.}
\end{figure}

\begin{figure*}[htbp]
\begin{minipage}[b]{0.45\linewidth}
\mbox{
\includegraphics[width=\textwidth]{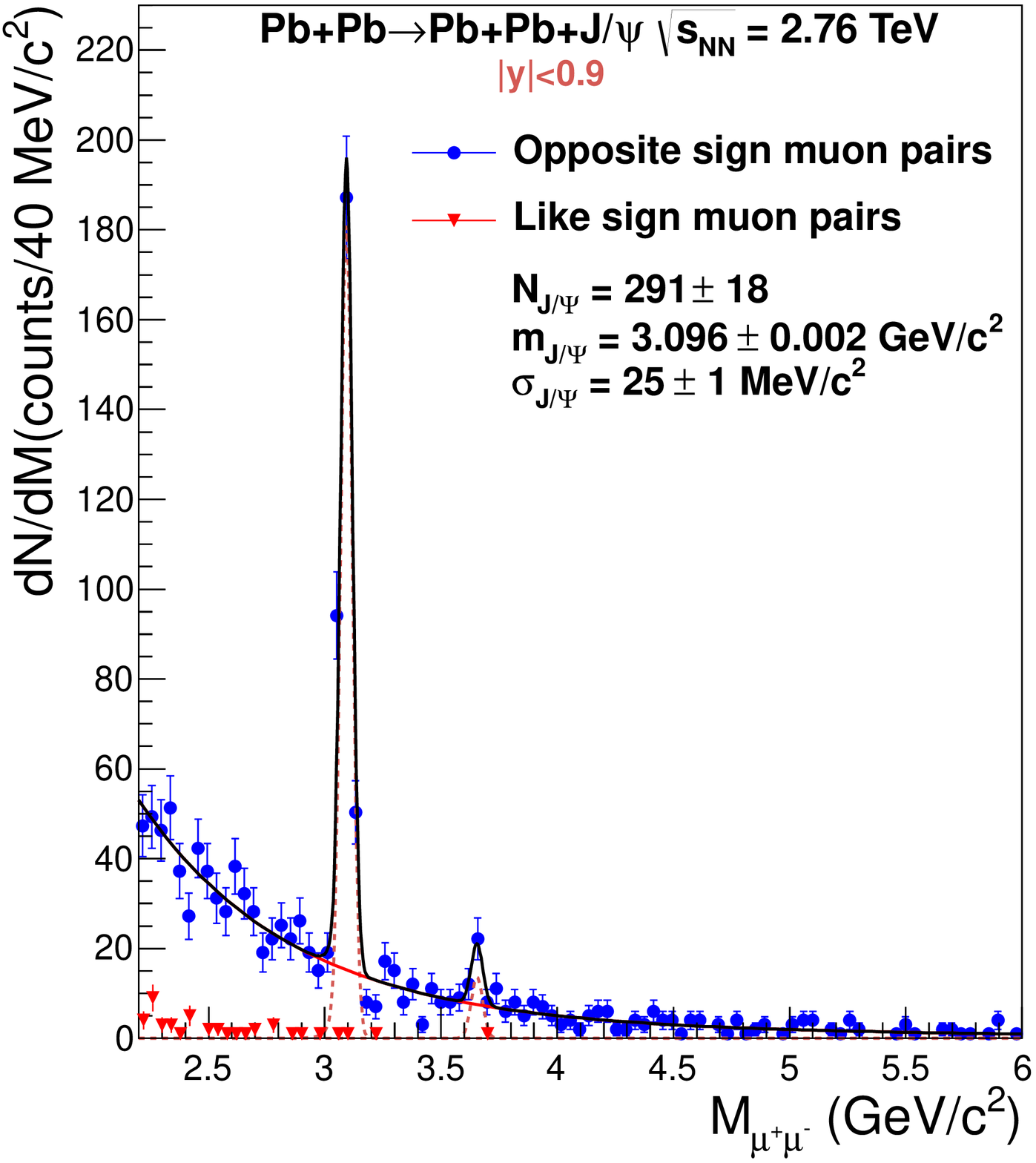}
\includegraphics[width=\textwidth]{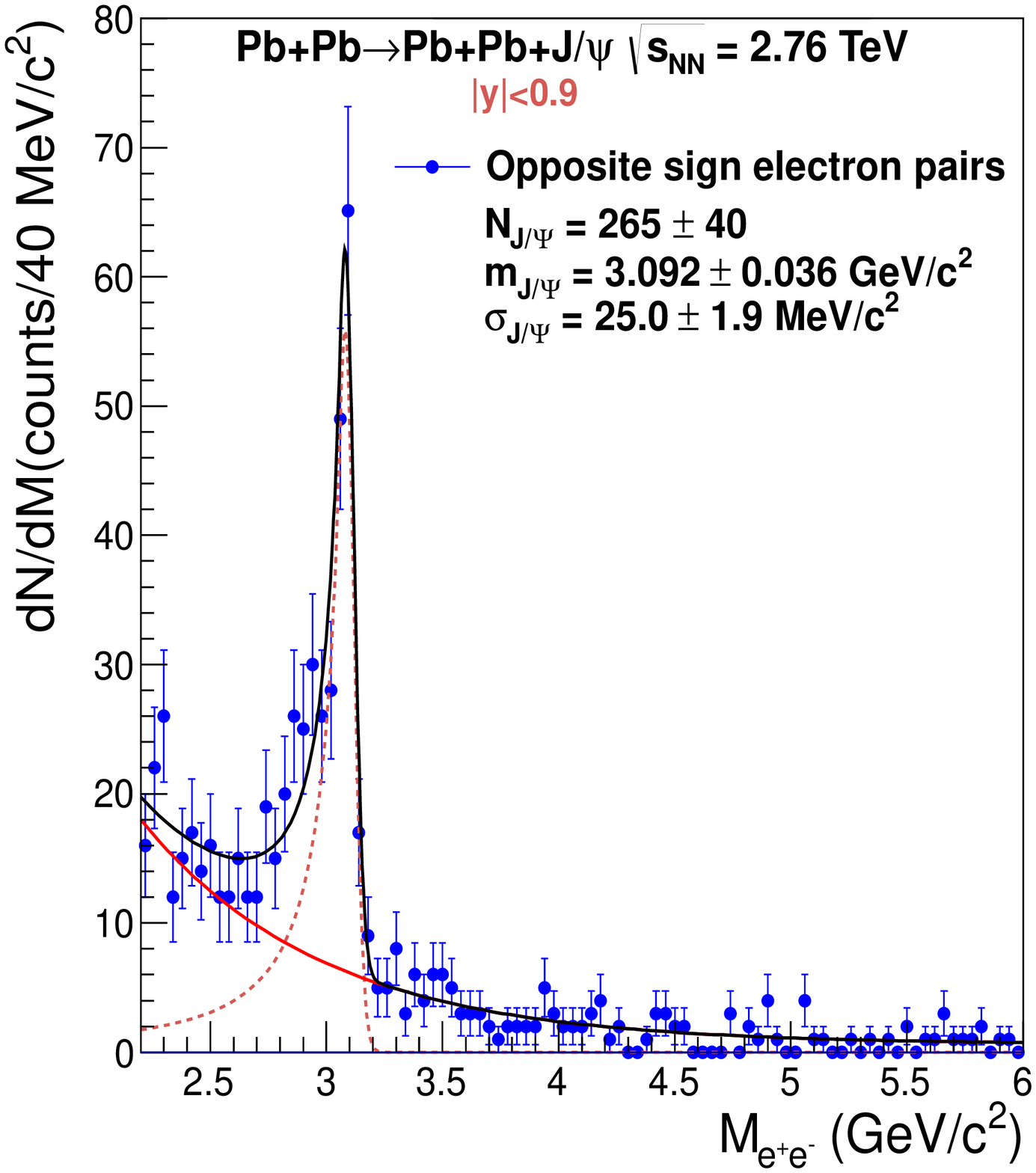}
}
\mbox{
\includegraphics[width=\textwidth]{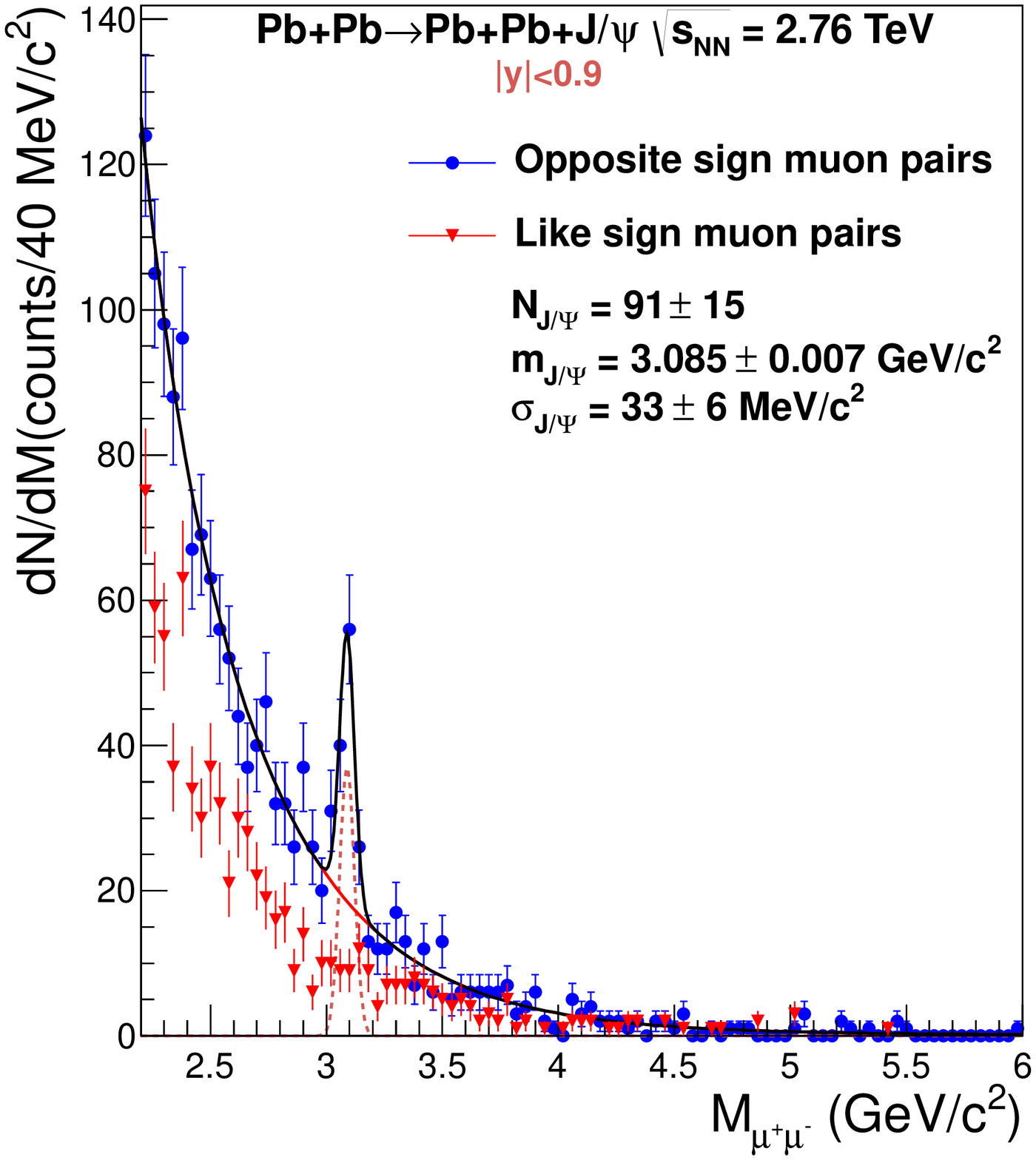}
\includegraphics[width=\textwidth]{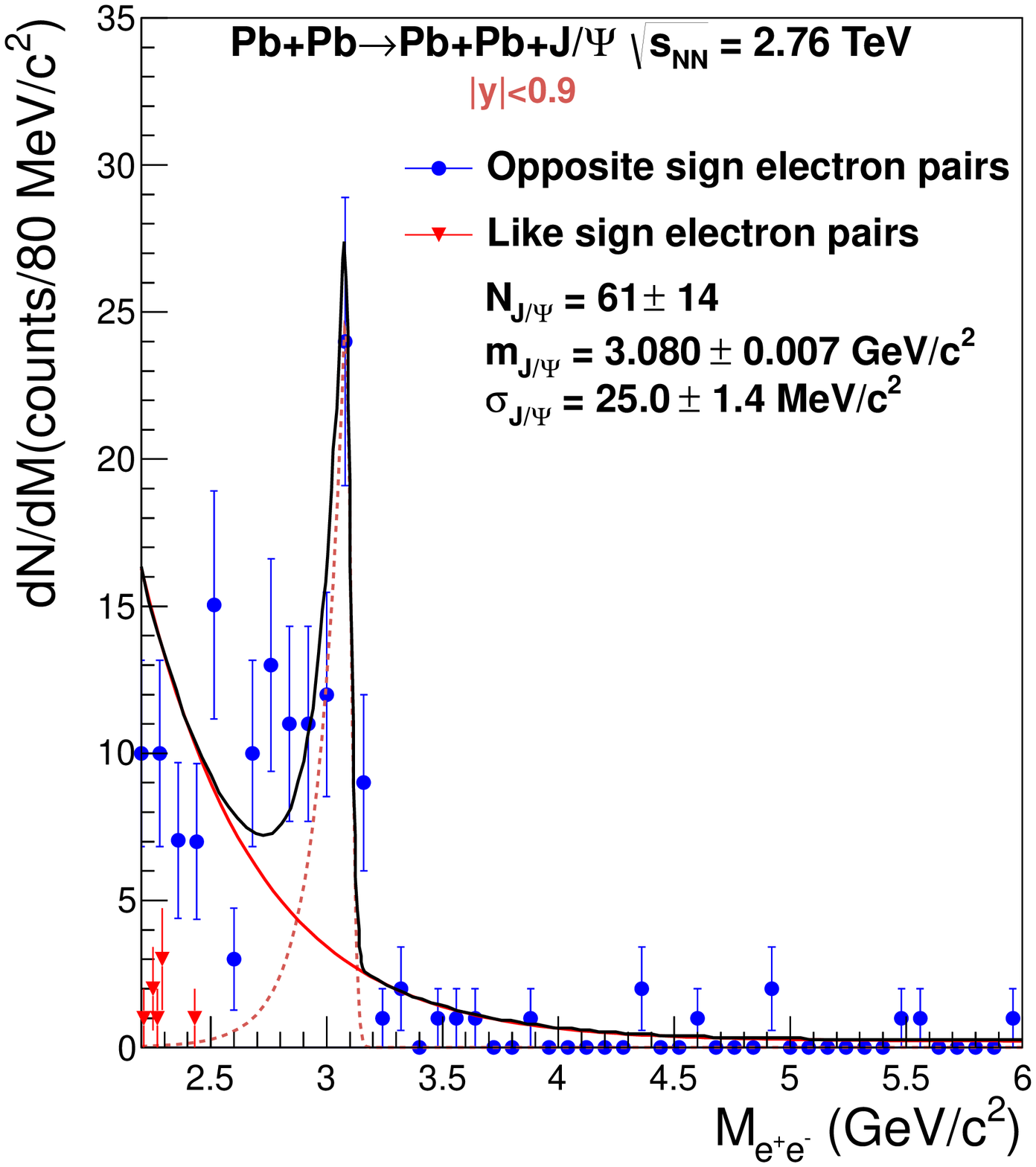}
}
\end{minipage}
\caption{Invariant mass distributions for ultra-peripheral Pb-Pb collisions at  $\sqrt{s_{\mathrm{NN}} } = 2.76$ TeV and -0.9~$<\it{y}<$~0.9 for 
events satisfying the event selection in Table~1, in the invariant mass range 
 2.2 $<$ \minv~$<$ 6~GeV/$c^{2}$. Coherent enriched sample~(top) and incoherent enriched sample~(bottom) for muons~(left) and electrons~(right).
Blue~(red) circles~(triangles) are opposite-sign~(like-sign) pairs. 
For like-sign pair the penultimate cut in Table~1 is replaced by the request of a same-sign pair.
No LS events were found for coherent di-electron events.}
\label{fig:yield}
\end{figure*}

\begin{figure}
\begin{center}
\includegraphics[width=1.\linewidth,keepaspectratio]{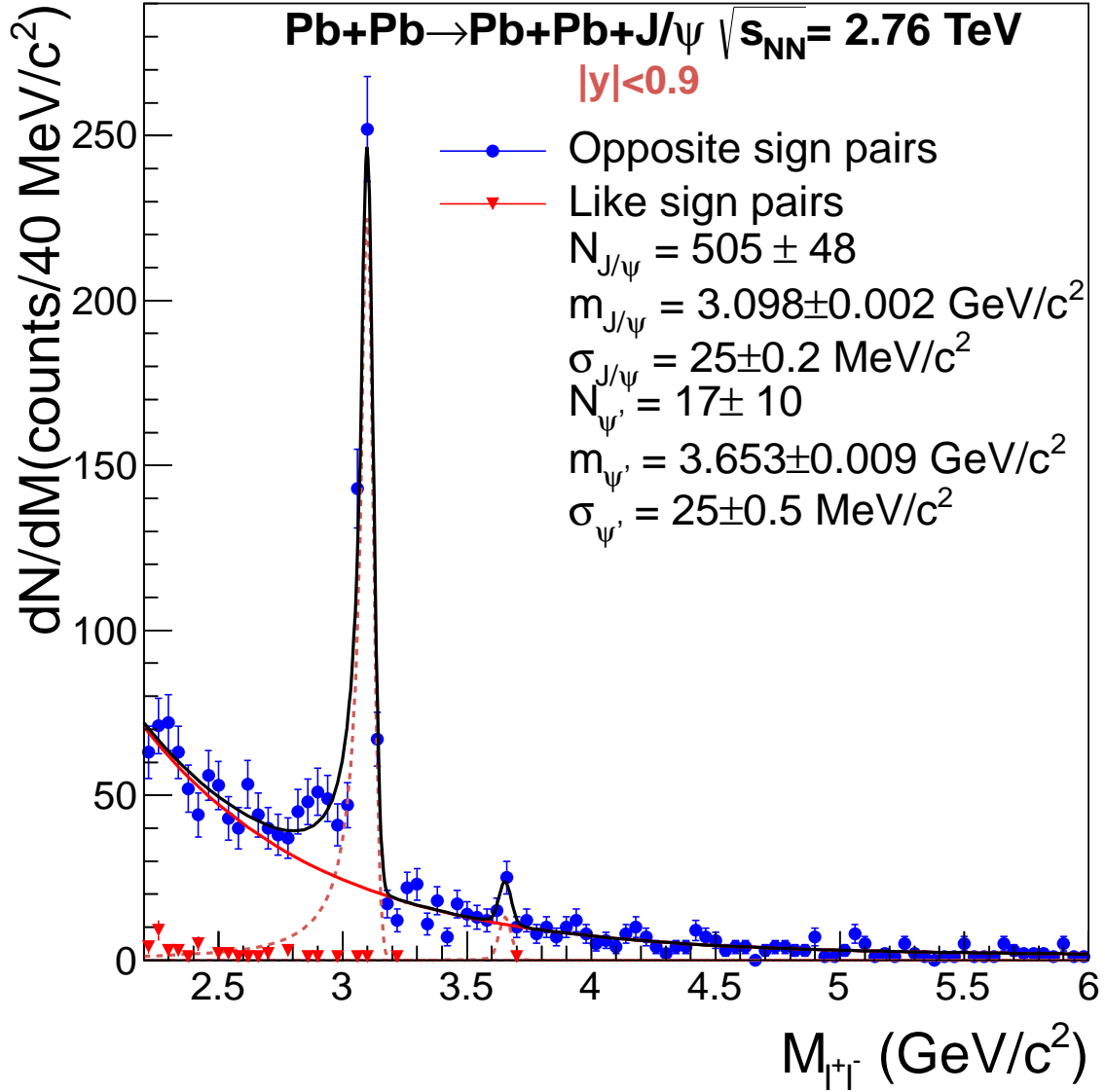}
\end{center}
\caption{\label{fig:all} Invariant mass distribution for
ultra-peripheral Pb-Pb collisions at $\sqrt{s_{\mathrm{NN}} } = 2.76$ TeV at -0.9$<$$\it{y}$$<$0.9
for events satisfying the event selection in Table~1, in
the invariant mass interval 2.2 $<$~\minv~$<$ 6~GeV/$c^{2}$. Coherent di-electron and di-muon candidates are summed together.
For like-sign pair the penultimate cut in Table~1 is replaced by the request of a same-sign pair.}
\end{figure}
\begin{figure*}[htbp]
\begin{minipage}[b]{0.45\linewidth}
\mbox{
\includegraphics[width=\textwidth]{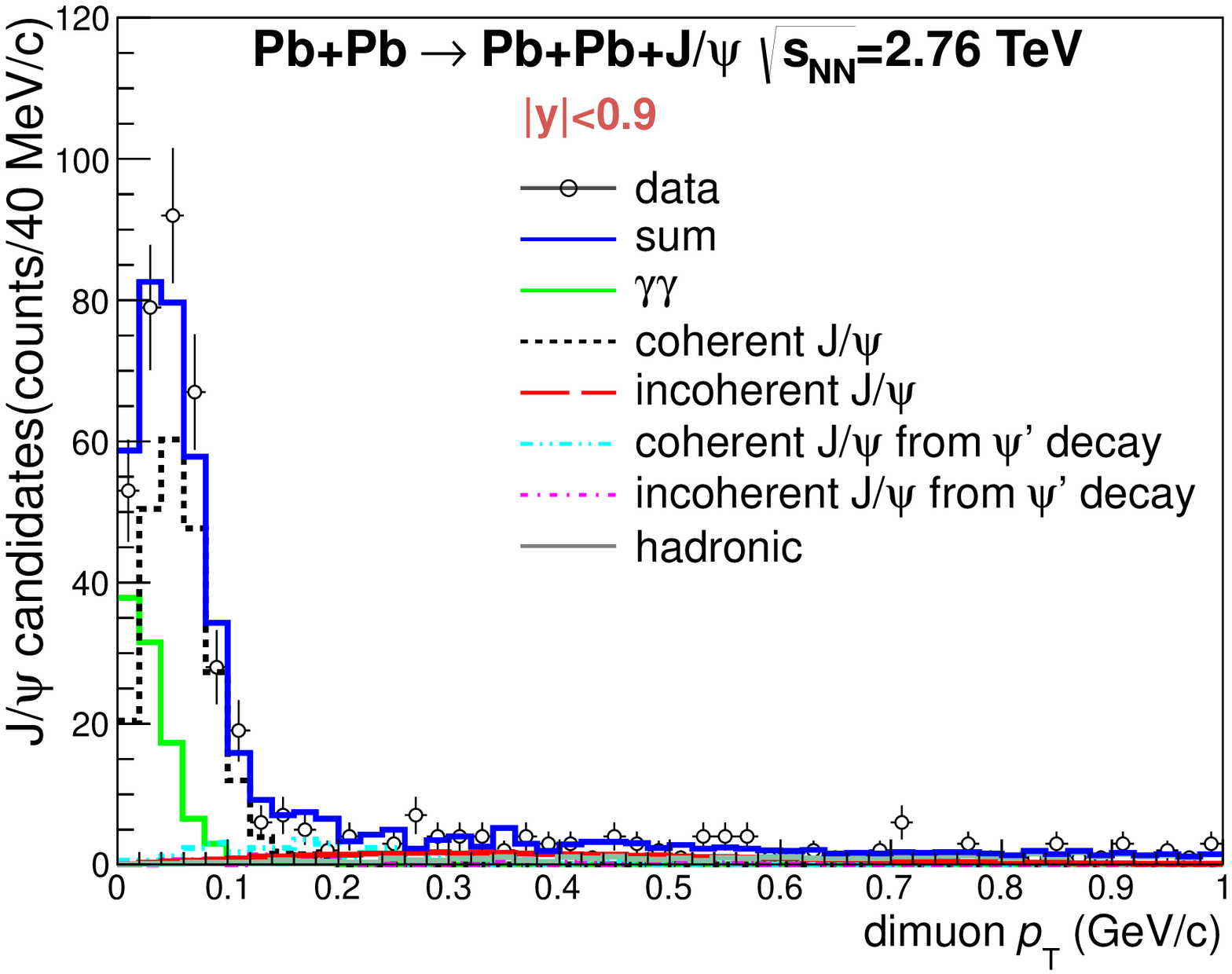}
\includegraphics[width=\textwidth]{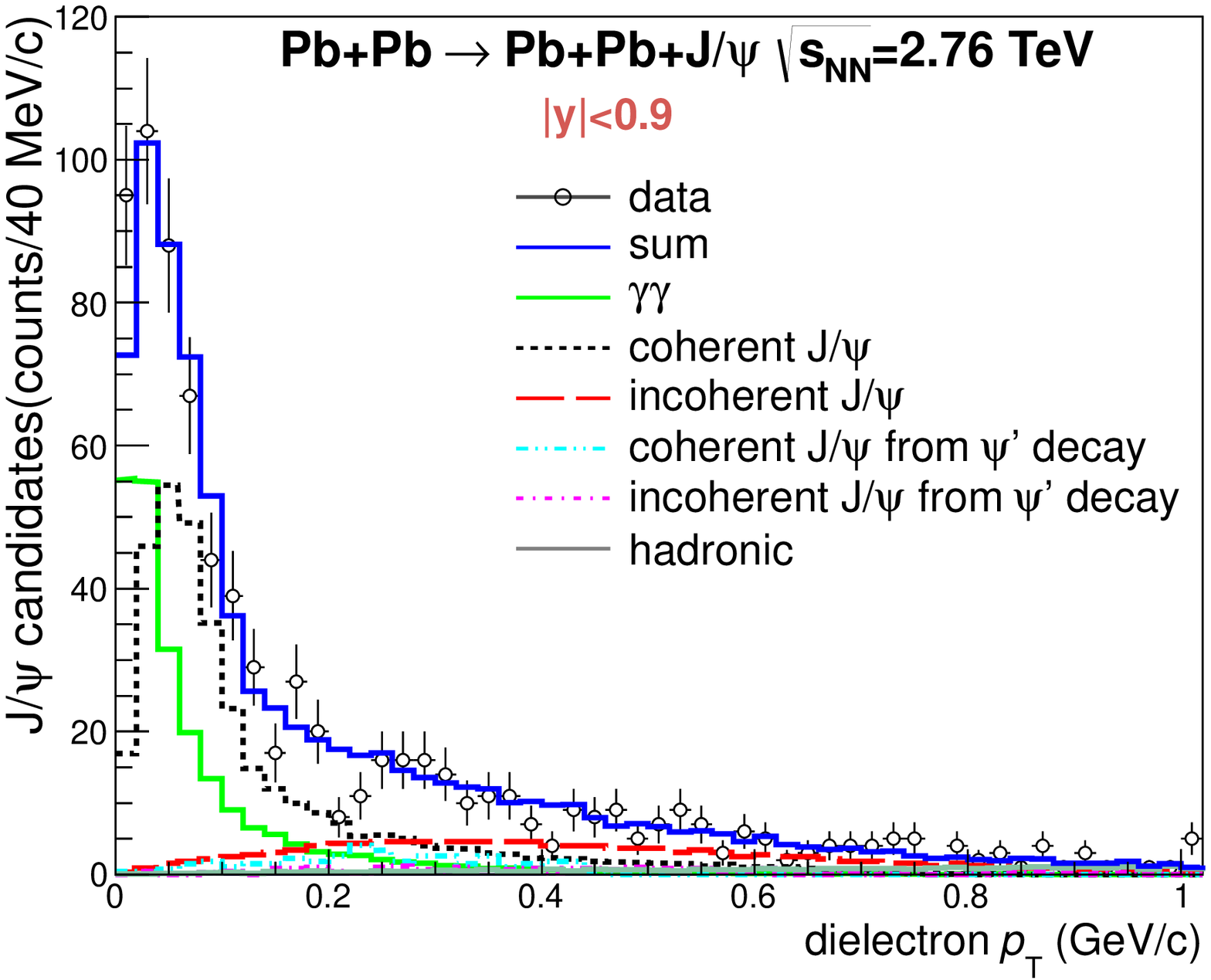}
}
\mbox{
\includegraphics[width=\textwidth]{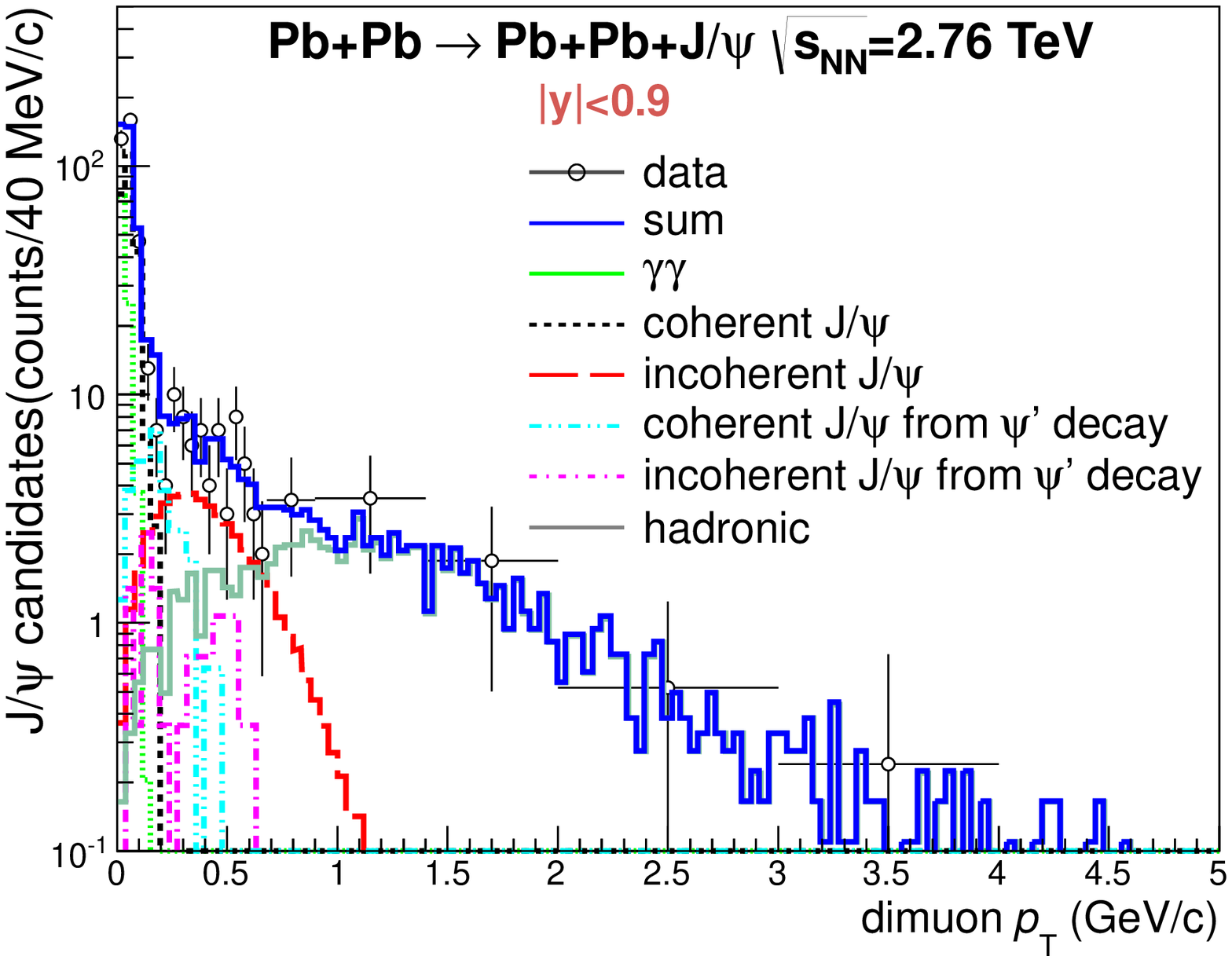}
\includegraphics[width=\textwidth]{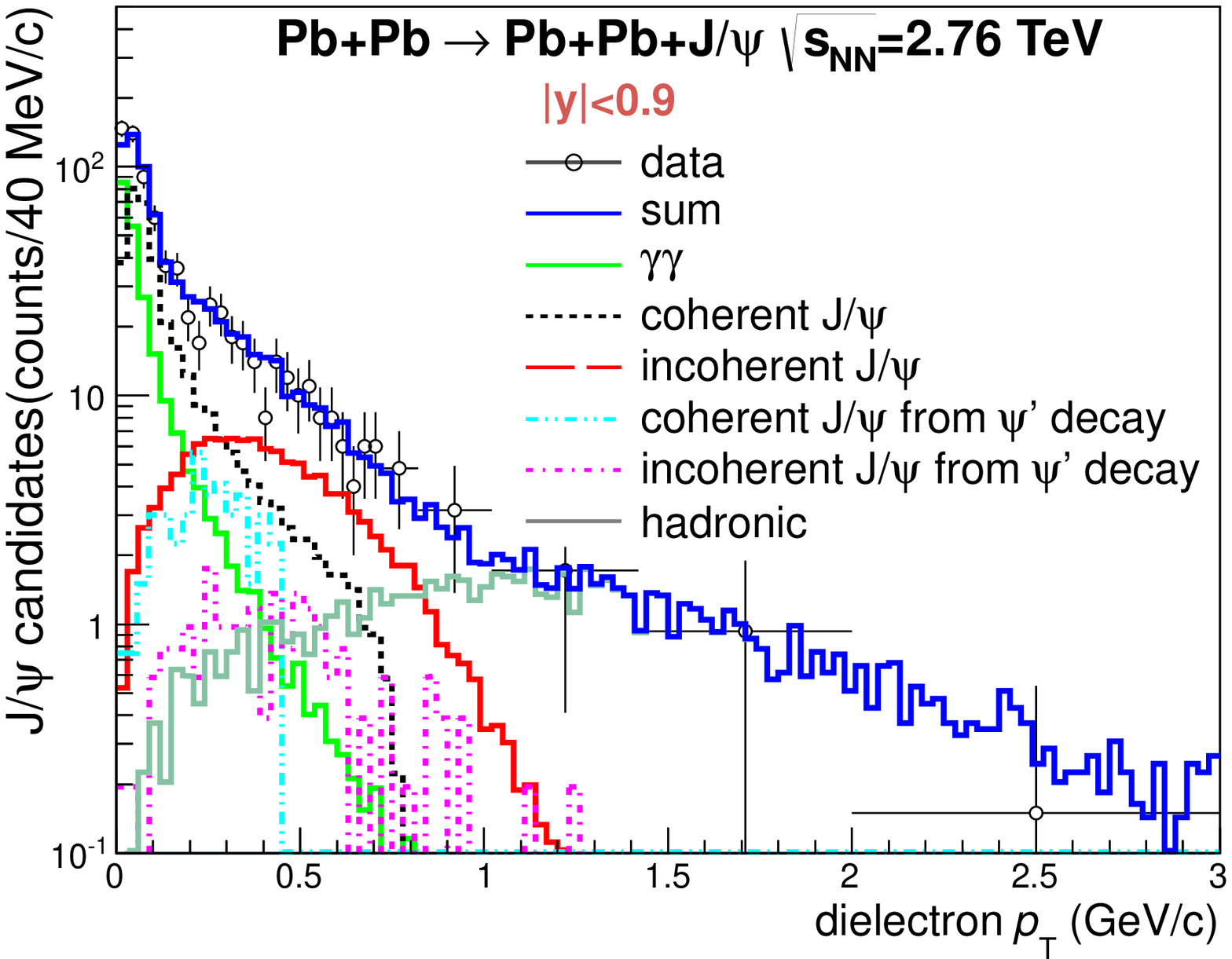}
}
\end{minipage}
\caption{\label{fig:pt}Di-muon~(left) and di-electron~(right) \pt~distribution for 
ultra-peripheral Pb-Pb collisions at $\sqrt{s_{\mathrm{NN}} } = 2.76$ TeV and -0.9~$<\it{y}<$~0.9 for
events satisfying the event selection 
in the invariant mass interval 3.0 $<$ \minv~$<$ 3.2~GeV/$c^{2}$ and 2.2 $<$ \minv~$<$ 3.2~GeV/$c^{2}$ respectively,
with the \pt-range extended to $p_{\rm T} <$~1~GeV/$c$~(top) and to $p_{\rm T} <$~5~GeV/$c$~(bottom). 
The data points are fitted summing
six different Monte Carlo templates: coherent \jpsi~production (black), incoherent \jpsi~production
(red), {\jpsi}s~from coherent $\psi^{'}$ decay (light blue), {\jpsi}s~from incoherent $\psi^{'}$ decay (violet),
$\gamma \gamma $~(green), and 
\jpsi~produced in peripheral hadronic collisions~(grey).
 The solid histogram (blue) is the sum.}
\end{figure*}
\begin{figure*}
\begin{center}
\includegraphics[width=1.\linewidth,keepaspectratio]{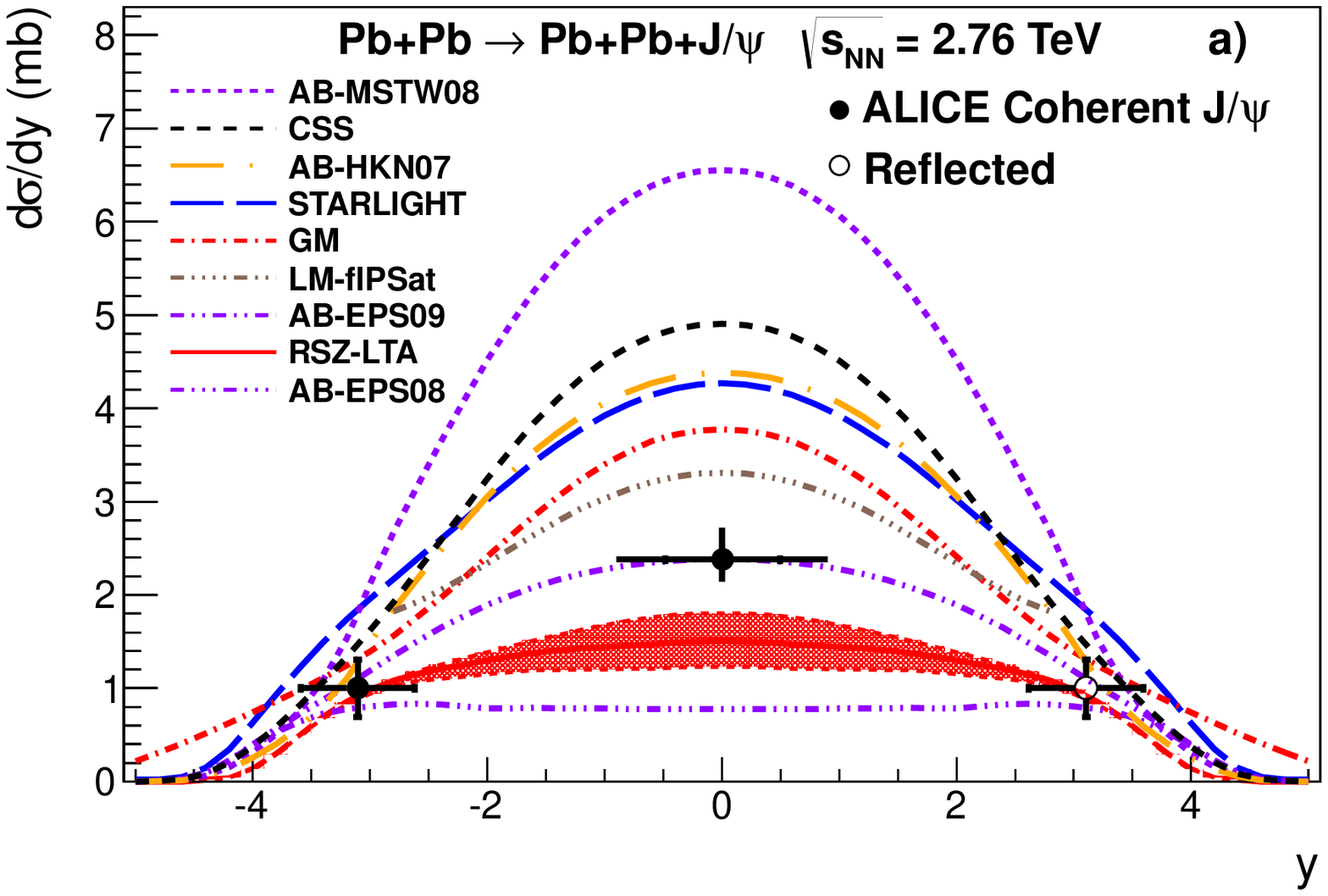}
\includegraphics[width=1.\linewidth,keepaspectratio]{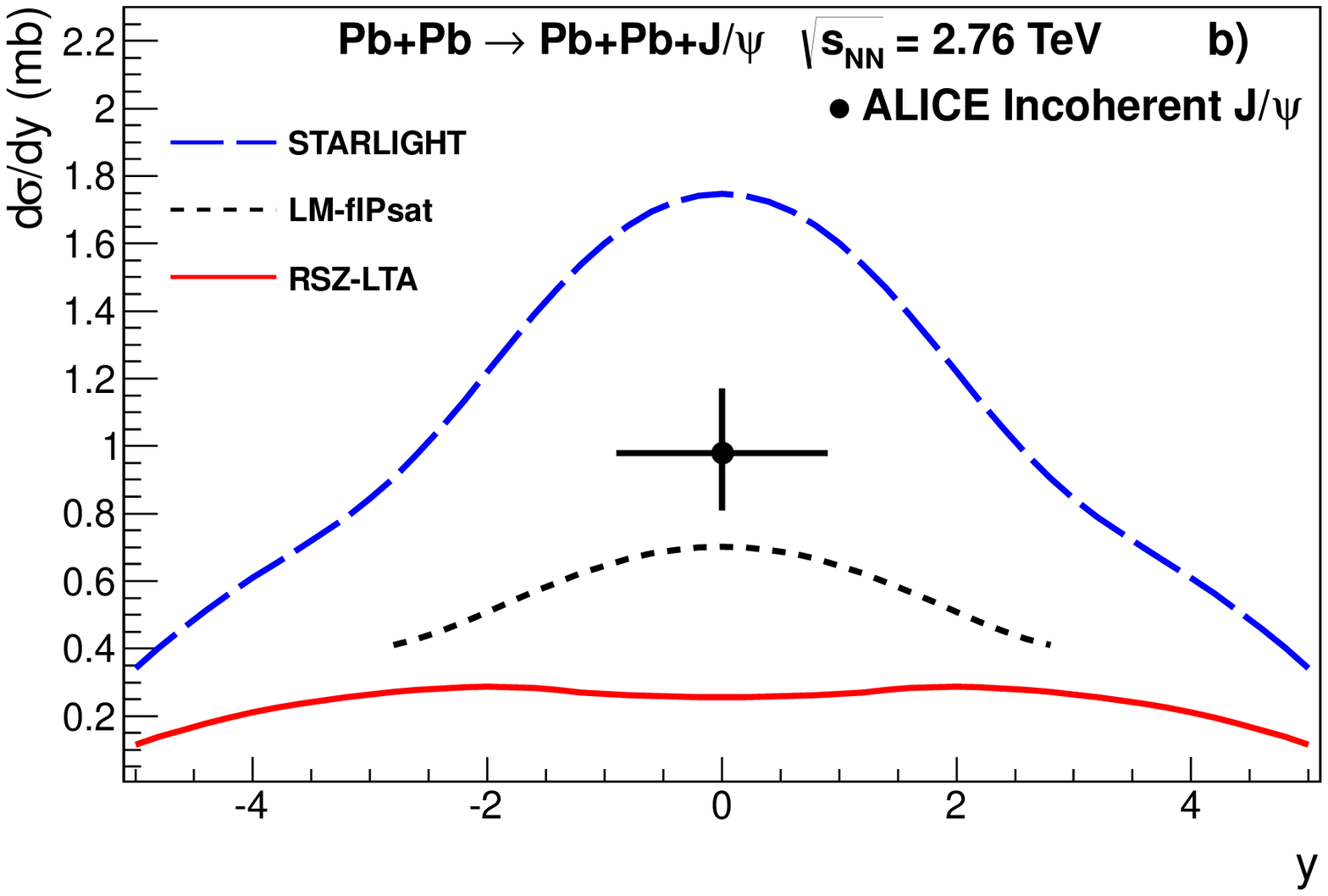}
\end{center}
\caption{\label{fig:theoryComparison} Measured differential cross section of \jpsi~photoproduction in
 ultra-peripheral Pb-Pb collisions at $\sqrt{s_{\mathrm{NN}} } = 2.76$ TeV 
at -0.9$<$$\it{y}$$<$0.9 for coherent a) 
 and incoherent b) events. The error is the quadratic sum of the
statistical and systematic errors. The theoretical calculations described in the text are also shown.}
\end{figure*}
\begin{figure*}
\begin{center}
\includegraphics[width=1.\linewidth,keepaspectratio]{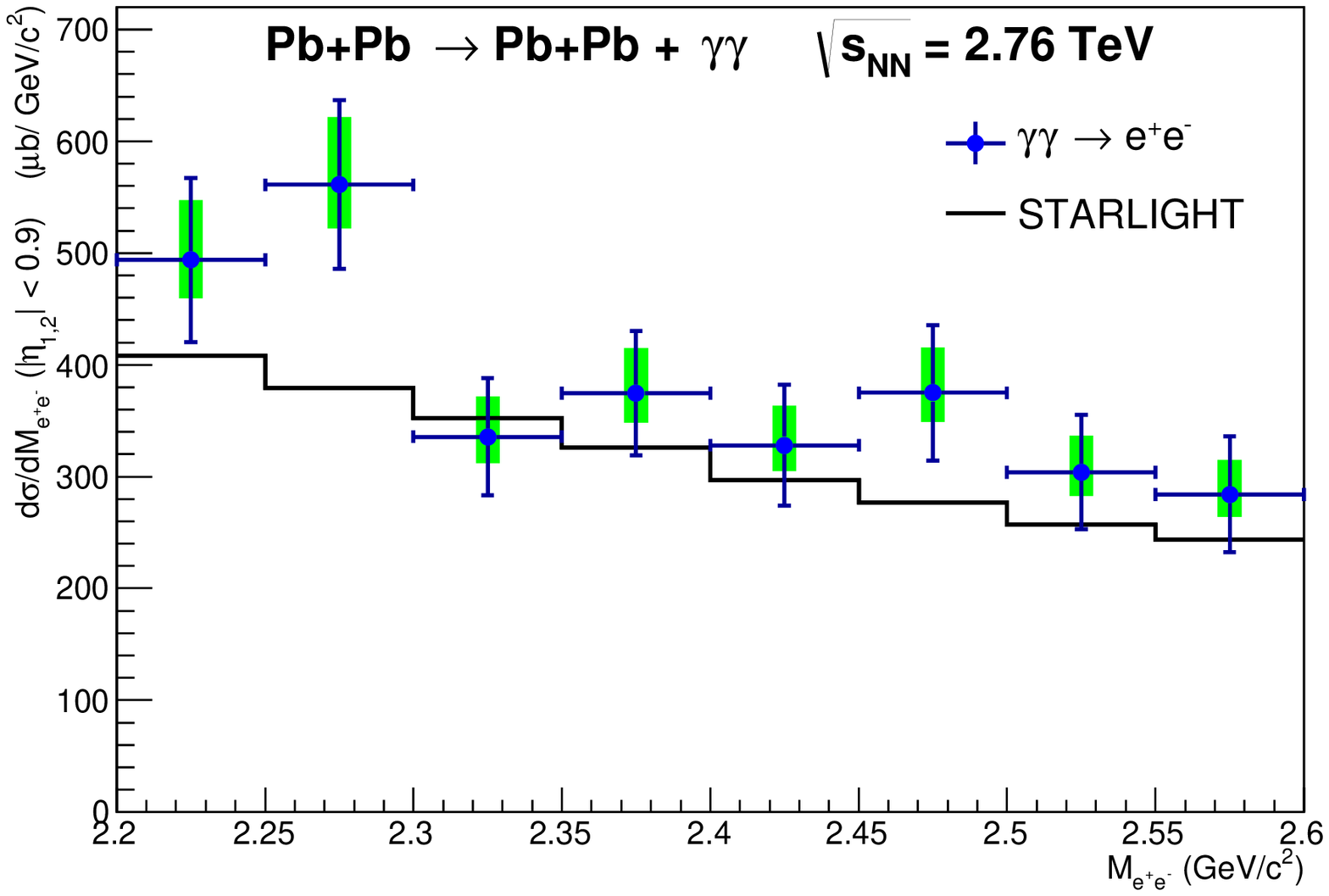}
\includegraphics[width=1.\linewidth,keepaspectratio]{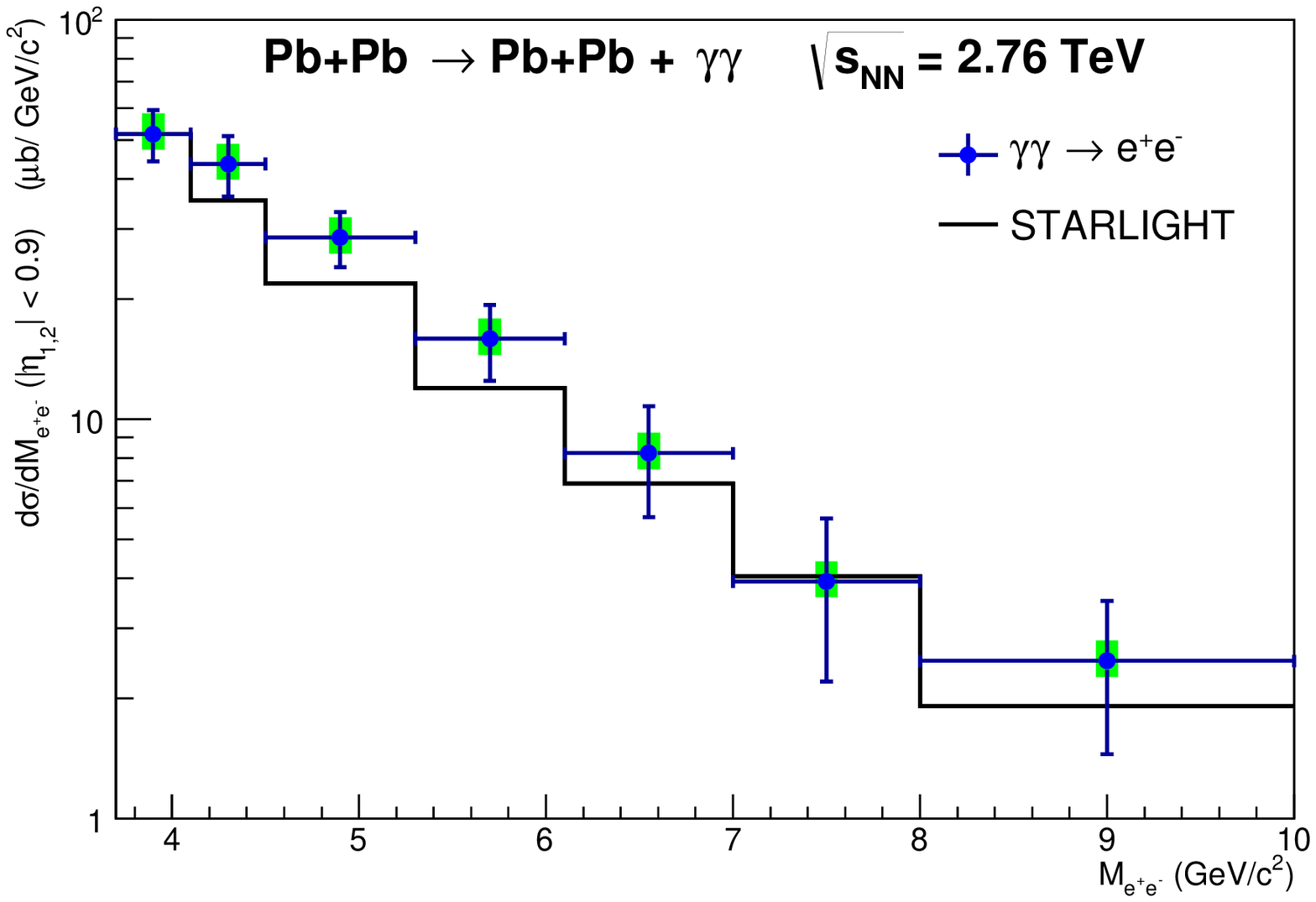}
\end{center}

\caption{\label{fig:gg} $\gamma\gamma \ \rightarrow e^+ e^-$ cross section~(blue circles) for
ultra-peripheral Pb-Pb collisions at $\sqrt{s_{\mathrm{NN}} } = 2.76$ TeV at -0.9$<$$\eta$$<$0.9
for events in the invariant mass interval 2.2 $<$ \minv~$<$ 2.6~GeV/$c^{2}$ (top) and 
3.7 $<$ \minv~$<$ 10~GeV/$c^{2}$~interval (bottom) compared to STARLIGHT simulation~(black line).
The blue(green) bars show the statistical (systematic) errors, respectively.}
\end{figure*}
\clearpage

\newenvironment{acknowledgement}{\relax}{\relax}
\begin{acknowledgement}
\section{Acknowledgements}
The ALICE collaboration would like to thank all its engineers and technicians for their invaluable contributions to the construction of the experiment and the CERN accelerator teams for the outstanding performance of the LHC complex.
\\
The ALICE collaboration acknowledges the following funding agencies for their support in building and
running the ALICE detector:
 \\
State Committee of Science,  World Federation of Scientists (WFS)
and Swiss Fonds Kidagan, Armenia,
 \\
Conselho Nacional de Desenvolvimento Cient\'{\i}fico e Tecnol\'{o}gico (CNPq), Financiadora de Estudos e Projetos (FINEP),
Funda\c{c}\~{a}o de Amparo \`{a} Pesquisa do Estado de S\~{a}o Paulo (FAPESP);
 \\
National Natural Science Foundation of China (NSFC), the Chinese Ministry of Education (CMOE)
and the Ministry of Science and Technology of China (MSTC);
 \\
Ministry of Education and Youth of the Czech Republic;
 \\
Danish Natural Science Research Council, the Carlsberg Foundation and the Danish National Research Foundation;
 \\
The European Research Council under the European Community's Seventh Framework Programme;
 \\
Helsinki Institute of Physics and the Academy of Finland;
 \\
French CNRS-IN2P3, the `Region Pays de Loire', `Region Alsace', `Region Auvergne' and CEA, France;
 \\
German BMBF and the Helmholtz Association;
\\
General Secretariat for Research and Technology, Ministry of
Development, Greece;
\\
Hungarian OTKA and National Office for Research and Technology (NKTH);
 \\
Department of Atomic Energy and Department of Science and Technology of the Government of India;
 \\
Istituto Nazionale di Fisica Nucleare (INFN) and Centro Fermi -
Museo Storico della Fisica e Centro Studi e Ricerche "Enrico
Fermi", Italy;
 \\
MEXT Grant-in-Aid for Specially Promoted Research, Ja\-pan;
 \\
Joint Institute for Nuclear Research, Dubna;
 \\
National Research Foundation of Korea (NRF);
 \\
CONACYT, DGAPA, M\'{e}xico, ALFA-EC and the EPLANET Program
(European Particle Physics Latin American Network)
 \\
Stichting voor Fundamenteel Onderzoek der Materie (FOM) and the Nederlandse Organisatie voor Wetenschappelijk Onderzoek (NWO), Netherlands;
 \\
Research Council of Norway (NFR);
 \\
Polish Ministry of Science and Higher Education;
 \\
National Authority for Scientific Research - NASR (Autoritatea Na\c{t}ional\u{a} pentru Cercetare \c{S}tiin\c{t}ific\u{a} - ANCS);
 \\
Ministry of Education and Science of Russian Federation, Russian
Academy of Sciences, Russian Federal Agency of Atomic Energy,
Russian Federal Agency for Science and Innovations and The Russian
Foundation for Basic Research;
 \\
Ministry of Education of Slovakia;
 \\
Department of Science and Technology, South Africa;
 \\
CIEMAT, EELA, Ministerio de Econom\'{i}a y Competitividad (MINECO) of Spain, Xunta de Galicia (Conseller\'{\i}a de Educaci\'{o}n),
CEA\-DEN, Cubaenerg\'{\i}a, Cuba, and IAEA (International Atomic Energy Agency);
 \\
Swedish Research Council (VR) and Knut $\&$ Alice Wallenberg
Foundation (KAW);
 \\
Ukraine Ministry of Education and Science;
 \\
United Kingdom Science and Technology Facilities Council (STFC);
 \\
The United States Department of Energy, the United States National
Science Foundation, the State of Texas, and the State of Ohio.
\end{acknowledgement}
\bibliographystyle{style}
\bibliography{biblio}

\newpage

\appendix
\section{The ALICE Collaboration}
\label{app:collab}

\begingroup
\small
\begin{flushleft}
E.~Abbas\Irefn{org36632}\And
B.~Abelev\Irefn{org1234}\And
J.~Adam\Irefn{org1274}\And
D.~Adamov\'{a}\Irefn{org1283}\And
A.M.~Adare\Irefn{org1260}\And
M.M.~Aggarwal\Irefn{org1157}\And
G.~Aglieri~Rinella\Irefn{org1192}\And
M.~Agnello\Irefn{org1313}\textsuperscript{,}\Irefn{org1017688}\And
A.G.~Agocs\Irefn{org1143}\And
A.~Agostinelli\Irefn{org1132}
% ...
\renewcommand\labelenumi{\textsuperscript{\theenumi}~}
\section*{Affiliation notes}
\renewcommand\theenumi{\roman{enumi}}
\begin{Authlist}
\item \Adef{M.V.Lomonosov Moscow State University, D.V.Skobeltsyn Institute of Nuclear Physics, Moscow, Russia}Also at: M.V.Lomonosov Moscow State University, D.V.Skobeltsyn Institute of Nuclear Physics, Moscow, Russia
\item \Adef{University of Belgrade, Faculty of Physics and "Vinvca" Institute of Nuclear Sciences, Belgrade, Serbia}Also at: University of Belgrade, Faculty of Physics and "Vinvca" Institute of Nuclear Sciences, Belgrade, Serbia
\item \Adef{Institute of Theoretical Physics, University of Wroclaw, Wroclaw, Poland}Also at: Institute of Theoretical Physics, University of Wroclaw, Wroclaw, Poland
\end{Authlist}
\section*{Collaboration Institutes}
\renewcommand\theenumi{\arabic{enumi}~}
\begin{Authlist}
\item \Idef{org36632}Academy of Scientific Research and Technology (ASRT), Cairo, Egypt
\item \Idef{org1332}A. I. Alikhanyan National Science Laboratory (Yerevan Physics Institute) Foundation, Yerevan, Armenia
\item \Idef{org1279}Benem\'{e}rita Universidad Aut\'{o}noma de Puebla, Puebla, Mexico
\item \Idef{org1220}Bogolyubov Institute for Theoretical Physics, Kiev, Ukraine
\item \Idef{org20959}Bose Institute, Department of Physics and Centre for Astroparticle Physics and Space Science (CAPSS), Kolkata, India
\item \Idef{org1262}Budker Institute for Nuclear Physics, Novosibirsk, Russia
\item \Idef{org1292}California Polytechnic State University, San Luis Obispo, California, United States
\item \Idef{org1329}Central China Normal University, Wuhan, China
\item \Idef{org14939}Centre de Calcul de l'IN2P3, Villeurbanne, France
\item \Idef{org1197}Centro de Aplicaciones Tecnol\'{o}gicas y Desarrollo Nuclear (CEADEN), Havana, Cuba
\item \Idef{org1242}Centro de Investigaciones Energ\'{e}ticas Medioambientales y Tecnol\'{o}gicas (CIEMAT), Madrid, Spain
\item \Idef{org1244}Centro de Investigaci\'{o}n y de Estudios Avanzados (CINVESTAV), Mexico City and M\'{e}rida, Mexico
\item \Idef{org1335}Centro Fermi - Museo Storico della Fisica e Centro Studi e Ricerche ``Enrico Fermi'', Rome, Italy
\item \Idef{org17347}Chicago State University, Chicago, United States
\item \Idef{org1288}Commissariat \`{a} l'Energie Atomique, IRFU, Saclay, France
\item \Idef{org15782}COMSATS Institute of Information Technology (CIIT), Islamabad, Pakistan
\item \Idef{org1294}Departamento de F\'{\i}sica de Part\'{\i}culas and IGFAE, Universidad de Santiago de Compostela, Santiago de Compostela, Spain
\item \Idef{org1106}Department of Physics Aligarh Muslim University, Aligarh, India
\item \Idef{org1121}Department of Physics and Technology, University of Bergen, Bergen, Norway
\item \Idef{org1162}Department of Physics, Ohio State University, Columbus, Ohio, United States
\item \Idef{org1300}Department of Physics, Sejong University, Seoul, South Korea
\item \Idef{org1268}Department of Physics, University of Oslo, Oslo, Norway
\item \Idef{org1315}Dipartimento di Fisica dell'Universit\`{a} and Sezione INFN, Trieste, Italy
\item \Idef{org1145}Dipartimento di Fisica dell'Universit\`{a} and Sezione INFN, Cagliari, Italy
\item \Idef{org1312}Dipartimento di Fisica dell'Universit\`{a} and Sezione INFN, Turin, Italy
\item \Idef{org1285}Dipartimento di Fisica dell'Universit\`{a} `La Sapienza' and Sezione INFN, Rome, Italy
\item \Idef{org1154}Dipartimento di Fisica e Astronomia dell'Universit\`{a} and Sezione INFN, Catania, Italy
\item \Idef{org1132}Dipartimento di Fisica e Astronomia dell'Universit\`{a} and Sezione INFN, Bologna, Italy
\item \Idef{org1270}Dipartimento di Fisica e Astronomia dell'Universit\`{a} and Sezione INFN, Padova, Italy
\item \Idef{org1290}Dipartimento di Fisica `E.R.~Caianiello' dell'Universit\`{a} and Gruppo Collegato INFN, Salerno, Italy
\item \Idef{org1103}Dipartimento di Scienze e Innovazione Tecnologica dell'Universit\`{a} del Piemonte Orientale and Gruppo Collegato INFN, Alessandria, Italy
\item \Idef{org1114}Dipartimento Interateneo di Fisica `M.~Merlin' and Sezione INFN, Bari, Italy
\item \Idef{org1237}Division of Experimental High Energy Physics, University of Lund, Lund, Sweden
\item \Idef{org1192}European Organization for Nuclear Research (CERN), Geneva, Switzerland
\item \Idef{org1227}Fachhochschule K\"{o}ln, K\"{o}ln, Germany
\item \Idef{org1122}Faculty of Engineering, Bergen University College, Bergen, Norway
\item \Idef{org1136}Faculty of Mathematics, Physics and Informatics, Comenius University, Bratislava, Slovakia
\item \Idef{org1274}Faculty of Nuclear Sciences and Physical Engineering, Czech Technical University in Prague, Prague, Czech Republic
\item \Idef{org1229}Faculty of Science, P.J.~\v{S}af\'{a}rik University, Ko\v{s}ice, Slovakia
\item \Idef{org1184}Frankfurt Institute for Advanced Studies, Johann Wolfgang Goethe-Universit\"{a}t Frankfurt, Frankfurt, Germany
\item \Idef{org1215}Gangneung-Wonju National University, Gangneung, South Korea
\item \Idef{org20958}Gauhati University, Department of Physics, Guwahati, India
\item \Idef{org1212}Helsinki Institute of Physics (HIP) and University of Jyv\"{a}skyl\"{a}, Jyv\"{a}skyl\"{a}, Finland
\item \Idef{org1203}Hiroshima University, Hiroshima, Japan
\item \Idef{org1254}Indian Institute of Technology Bombay (IIT), Mumbai, India
\item \Idef{org36378}Indian Institute of Technology Indore, Indore, India (IITI)
\item \Idef{org1266}Institut de Physique Nucl\'{e}aire d'Orsay (IPNO), Universit\'{e} Paris-Sud, CNRS-IN2P3, Orsay, France
\item \Idef{org1277}Institute for High Energy Physics, Protvino, Russia
\item \Idef{org1249}Institute for Nuclear Research, Academy of Sciences, Moscow, Russia
\item \Idef{org1320}Nikhef, National Institute for Subatomic Physics and Institute for Subatomic Physics of Utrecht University, Utrecht, Netherlands
\item \Idef{org1250}Institute for Theoretical and Experimental Physics, Moscow, Russia
\item \Idef{org1230}Institute of Experimental Physics, Slovak Academy of Sciences, Ko\v{s}ice, Slovakia
\item \Idef{org1127}Institute of Physics, Bhubaneswar, India
\item \Idef{org1275}Institute of Physics, Academy of Sciences of the Czech Republic, Prague, Czech Republic
\item \Idef{org1139}Institute of Space Sciences (ISS), Bucharest, Romania
\item \Idef{org27399}Institut f\"{u}r Informatik, Johann Wolfgang Goethe-Universit\"{a}t Frankfurt, Frankfurt, Germany
\item \Idef{org1185}Institut f\"{u}r Kernphysik, Johann Wolfgang Goethe-Universit\"{a}t Frankfurt, Frankfurt, Germany
\item \Idef{org1177}Institut f\"{u}r Kernphysik, Technische Universit\"{a}t Darmstadt, Darmstadt, Germany
\item \Idef{org1256}Institut f\"{u}r Kernphysik, Westf\"{a}lische Wilhelms-Universit\"{a}t M\"{u}nster, M\"{u}nster, Germany
\item \Idef{org1246}Instituto de Ciencias Nucleares, Universidad Nacional Aut\'{o}noma de M\'{e}xico, Mexico City, Mexico
\item \Idef{org1247}Instituto de F\'{\i}sica, Universidad Nacional Aut\'{o}noma de M\'{e}xico, Mexico City, Mexico
\item \Idef{org1308}Institut Pluridisciplinaire Hubert Curien (IPHC), Universit\'{e} de Strasbourg, CNRS-IN2P3, Strasbourg, France
\item \Idef{org1182}Joint Institute for Nuclear Research (JINR), Dubna, Russia
\item \Idef{org1199}Kirchhoff-Institut f\"{u}r Physik, Ruprecht-Karls-Universit\"{a}t Heidelberg, Heidelberg, Germany
\item \Idef{org20954}Korea Institute of Science and Technology Information, Daejeon, South Korea
\item \Idef{org1017642}KTO Karatay University, Konya, Turkey
\item \Idef{org1160}Laboratoire de Physique Corpusculaire (LPC), Clermont Universit\'{e}, Universit\'{e} Blaise Pascal, CNRS--IN2P3, Clermont-Ferrand, France
\item \Idef{org1194}Laboratoire de Physique Subatomique et de Cosmologie (LPSC), Universit\'{e} Joseph Fourier, CNRS-IN2P3, Institut Polytechnique de Grenoble, Grenoble, France
\item \Idef{org1187}Laboratori Nazionali di Frascati, INFN, Frascati, Italy
\item \Idef{org1232}Laboratori Nazionali di Legnaro, INFN, Legnaro, Italy
\item \Idef{org1125}Lawrence Berkeley National Laboratory, Berkeley, California, United States
\item \Idef{org1234}Lawrence Livermore National Laboratory, Livermore, California, United States
\item \Idef{org1251}Moscow Engineering Physics Institute, Moscow, Russia
\item \Idef{org1322}National Centre for Nuclear Studies, Warsaw, Poland
\item \Idef{org1140}National Institute for Physics and Nuclear Engineering, Bucharest, Romania
\item \Idef{org1017626}National Institute of Science Education and Research, Bhubaneswar, India
\item \Idef{org1165}Niels Bohr Institute, University of Copenhagen, Copenhagen, Denmark
\item \Idef{org1109}Nikhef, National Institute for Subatomic Physics, Amsterdam, Netherlands
\item \Idef{org1283}Nuclear Physics Institute, Academy of Sciences of the Czech Republic, \v{R}e\v{z} u Prahy, Czech Republic
\item \Idef{org1264}Oak Ridge National Laboratory, Oak Ridge, Tennessee, United States
\item \Idef{org1189}Petersburg Nuclear Physics Institute, Gatchina, Russia
\item \Idef{org1170}Physics Department, Creighton University, Omaha, Nebraska, United States
\item \Idef{org1157}Physics Department, Panjab University, Chandigarh, India
\item \Idef{org1112}Physics Department, University of Athens, Athens, Greece
\item \Idef{org1152}Physics Department, University of Cape Town and  iThemba LABS, National Research Foundation, Somerset West, South Africa
\item \Idef{org1209}Physics Department, University of Jammu, Jammu, India
\item \Idef{org1207}Physics Department, University of Rajasthan, Jaipur, India
\item \Idef{org1200}Physikalisches Institut, Ruprecht-Karls-Universit\"{a}t Heidelberg, Heidelberg, Germany
\item \Idef{org1017688}Politecnico di Torino, Turin, Italy
\item \Idef{org1325}Purdue University, West Lafayette, Indiana, United States
\item \Idef{org1281}Pusan National University, Pusan, South Korea
\item \Idef{org1176}Research Division and ExtreMe Matter Institute EMMI, GSI Helmholtzzentrum f\"ur Schwerionenforschung, Darmstadt, Germany
\item \Idef{org1334}Rudjer Bo\v{s}kovi\'{c} Institute, Zagreb, Croatia
\item \Idef{org1298}Russian Federal Nuclear Center (VNIIEF), Sarov, Russia
\item \Idef{org1252}Russian Research Centre Kurchatov Institute, Moscow, Russia
\item \Idef{org1224}Saha Institute of Nuclear Physics, Kolkata, India
\item \Idef{org1130}School of Physics and Astronomy, University of Birmingham, Birmingham, United Kingdom
\item \Idef{org1338}Secci\'{o}n F\'{\i}sica, Departamento de Ciencias, Pontificia Universidad Cat\'{o}lica del Per\'{u}, Lima, Peru
\item \Idef{org1155}Sezione INFN, Catania, Italy
\item \Idef{org1313}Sezione INFN, Turin, Italy
\item \Idef{org1271}Sezione INFN, Padova, Italy
\item \Idef{org1133}Sezione INFN, Bologna, Italy
\item \Idef{org1146}Sezione INFN, Cagliari, Italy
\item \Idef{org1316}Sezione INFN, Trieste, Italy
\item \Idef{org1115}Sezione INFN, Bari, Italy
\item \Idef{org1286}Sezione INFN, Rome, Italy
\item \Idef{org36377}Nuclear Physics Group, STFC Daresbury Laboratory, Daresbury, United Kingdom
\item \Idef{org1258}SUBATECH, Ecole des Mines de Nantes, Universit\'{e} de Nantes, CNRS-IN2P3, Nantes, France
\item \Idef{org35706}Suranaree University of Technology, Nakhon Ratchasima, Thailand
\item \Idef{org1304}Technical University of Split FESB, Split, Croatia
\item \Idef{org1017659}Technische Universit\"{a}t M\"{u}nchen, Munich, Germany
\item \Idef{org1168}The Henryk Niewodniczanski Institute of Nuclear Physics, Polish Academy of Sciences, Cracow, Poland
\item \Idef{org17361}The University of Texas at Austin, Physics Department, Austin, TX, United States
\item \Idef{org1173}Universidad Aut\'{o}noma de Sinaloa, Culiac\'{a}n, Mexico
\item \Idef{org1296}Universidade de S\~{a}o Paulo (USP), S\~{a}o Paulo, Brazil
\item \Idef{org1149}Universidade Estadual de Campinas (UNICAMP), Campinas, Brazil
\item \Idef{org1239}Universit\'{e} de Lyon, Universit\'{e} Lyon 1, CNRS/IN2P3, IPN-Lyon, Villeurbanne, France
\item \Idef{org1205}University of Houston, Houston, Texas, United States
\item \Idef{org20371}University of Technology and Austrian Academy of Sciences, Vienna, Austria
\item \Idef{org1222}University of Tennessee, Knoxville, Tennessee, United States
\item \Idef{org1310}University of Tokyo, Tokyo, Japan
\item \Idef{org1318}University of Tsukuba, Tsukuba, Japan
\item \Idef{org21360}Eberhard Karls Universit\"{a}t T\"{u}bingen, T\"{u}bingen, Germany
\item \Idef{org1225}Variable Energy Cyclotron Centre, Kolkata, India
\item \Idef{org1017687}Vestfold University College, Tonsberg, Norway
\item \Idef{org1306}V.~Fock Institute for Physics, St. Petersburg State University, St. Petersburg, Russia
\item \Idef{org1323}Warsaw University of Technology, Warsaw, Poland
\item \Idef{org1179}Wayne State University, Detroit, Michigan, United States
\item \Idef{org1143}Wigner Research Centre for Physics, Hungarian Academy of Sciences, Budapest, Hungary
\item \Idef{org1260}Yale University, New Haven, Connecticut, United States
\item \Idef{org15649}Yildiz Technical University, Istanbul, Turkey
\item \Idef{org1301}Yonsei University, Seoul, South Korea
\item \Idef{org1327}Zentrum f\"{u}r Technologietransfer und Telekommunikation (ZTT), Fachhochschule Worms, Worms, Germany
\end{Authlist}
\endgroup

%
%

\end{document}